\documentclass[a4paper,fleqn,longmktitle]{cas-sc}

\usepackage[T1]{fontenc}
\usepackage[utf8]{inputenc}

\makeatletter
\def\@seccntformat#1{\@ifundefined{#1@cntformat}%
   {\csname the#1\endcsname\quad}
   {\csname #1@cntformat\endcsname}
}
\makeatother

\usepackage[numbers]{natbib}
\usepackage{rotating}
\usepackage{gensymb}
\usepackage[section]{placeins}
\usepackage{nomencl} 
\usepackage{framed} 
\usepackage{multicol} 
\usepackage{tabularx}
\usepackage{pdflscape}
\usepackage{longtable}

\usepackage{nomencl} 
\usepackage[boxruled]{algorithm2e} 
\makenomenclature
\renewcommand*\nompreamble{\begin{multicols}{2}}
\renewcommand*\nompostamble{\end{multicols}}

\renewcommand\nomgroup[1]{%
  \item[\bfseries
  \ifstrequal{#1}{A}{Abbreviations}{%
  \ifstrequal{#1}{V}{Variables}{%
  \ifstrequal{#1}{P}{Parameter}{%
  \ifstrequal{#1}{S}{Sets}}}}
]}

\begin{document}
\let\WriteBookmarks\relax
\def\floatpagepagefraction{1}
\def\textpagefraction{.001}
\shorttitle{Flexible electrification}
\shortauthors{Göke et al.}

\title [mode = title]{How flexible electrification can integrate fluctuating renewables}                  

\author[1,2]{Leonard Göke}[orcid=0000-0002-3219-7587]
\cormark[1]
\ead{lgo@wip.tu-berlin.de}
\author[3,1]{Jens Weibezahn}[orcid=0000-0003-1202-1709]
\author[1,2]{Mario Kendziorski}[orcid=0000-0001-5508-2953]

\address[1]{Workgroup for Infrastructure Policy (WIP), Technische Universität Berlin,Straße des 17.\,Juni 135, 10623 Berlin, Germany}
 \address[2]{Energy, Transportation, Environment Department, German Institute for Economic Research (DIW Berlin),Mohrenstraße 58, 10117 Berlin,Germany}
 \address[3]{Copenhagen School of Energy Infrastructure (CSEI), Department of Economics, Copenhagen Business School,Porcelænshaven 16A,2000 Frederiksberg,Denmark}

\cortext[cor1]{Corresponding author: \href{mailto:lgo@wip.tu-berlin.de}{lgo@wip.tu-berlin.de}}

\begin{abstract}
To phase-out fossil fuels, energy systems must shift to renewable electricity as the main source of primary energy. In this paper, we analyze how electrification can support the integration of fluctuating renewables, like wind and PV, and mitigate the need for storage and thermal backup plants. Using a cost minimizing model for system planning, we find substantial benefits of electricity demand in heating, transport, and industry adapting to supply. In Germany, flexible demand halves the residual peak-load and the residual demand and reduces excess generation by 80\%. Flexible operation of electrolyzers has the most significant impact accounting for 42\% of the reduction in residual peak-load and 59\% in residual demand. District heating networks and BEVs also provide substantial flexibility, while the contribution of space and process heating is negligible. The results are robust to restrictions on the expansion of the transmission grid.
\end{abstract}

\begin{keywords}
macro-energy systems \sep sector integration \sep decarbonization \sep flexibility \sep integrated energy system \sep flexible electricity demand
\end{keywords}

\maketitle

\begin{table*}[pos=!t]
   \begin{framed}
     \printnomenclature[1cm]
   \end{framed}
\end{table*}

\nomenclature[A, 01]{BEV}{battery electric vehicle}
\nomenclature[A, 02]{CC}{combined cycle}
\nomenclature[A, 03]{CHP}{combined heat and power}
\nomenclature[A, 04]{HVAC}{high-voltage alternating current}
\nomenclature[A, 05]{HVDC}{high-voltage direct current}
\nomenclature[A, 06]{NTC}{net-transfer capacity}
\nomenclature[A, 07]{PtX}{power-to-x}
\nomenclature[A, 08]{PV}{photovoltaic}
\nomenclature[A, 09]{OC}{open cycle}

\nomenclature[V, 01]{$K^{gen}_{r,i}$}{generation capacity}
\nomenclature[V, 02]{$K^{st}_{r,i}$}{storage power capacity}
\nomenclature[V, 03]{$K^{lvl}_{r,i}$}{storage energy capacity}
\nomenclature[V, 04]{$K^{exc}_{r,r',i}$}{transmission capacity}
\nomenclature[V, 05]{$G_{t,r,i,c}$}{generation quantity}
\nomenclature[V, 06]{$U_{t,r,i,c}$}{use quantity}
\nomenclature[V, 07]{$S_{t,r,i,c}^{in}$}{charged quantity}
\nomenclature[V, 08]{$S_{t,r,i,c}^{out}$}{discharged quantity}
\nomenclature[V, 09]{$S_{t,r,i}^{lvl}$}{storage level}
\nomenclature[V, 10]{$E_{t,r,r',i}$}{exchange quantity}

\nomenclature[P, 01]{$v_{t,r,i,c}^{var}$}{variable costs of generation}
\nomenclature[P, 02]{$v_{t,r,i,c}^{fix}$}{fixed costs of storage or generation capacity}
\nomenclature[P, 03]{$v_{r,r',c}^{fix}$}{fixed costs of transmission capacity}
\nomenclature[P, 04]{$d_{t,r,c}$}{demand}
\nomenclature[P, 05]{$\alpha_{t,r,i}$}{capacity factor}
\nomenclature[P, 06]{$\eta_{t,r,i}$}{conversion efficiency}
\nomenclature[P, 07]{$\delta_{t,r,i}$}{self-discharge rate}
\nomenclature[P, 08]{$\rho_{t,r,i}$}{charging efficiency}
\nomenclature[P, 09]{$p$}{peak demand}

\nomenclature[S, 01]{$t$}{time-steps}
\nomenclature[S, 02]{$r$}{regions}
\nomenclature[S, 03]{$i$}{technologies}
\nomenclature[S, 04]{$c$}{energy carriers}

\section{Introduction}  \label{intro}

International governments are pursuing different strategies to combat climate change and keep global warming \textit{"well below 2 degrees [...] compared to pre-industrial levels"}, as stated in the Paris Climate Agreement \citep{Paris2015}. Yet, policies have two common denominators: First, expanding electricity generation from wind or photovoltaic (PV), and second, utilizing more electricity in the heating, transport, or industry sector.

Electricity generation from wind and PV is for instance at the heart of the European Union's energy policy, part of the Inflation Reduction Act by the US government, and a key element of the Chinese energy strategy \citep{ec2022, wh2022, reuters2022b}. Wind and solar offer a great technical potential, exceeding global primary consumption at least three times, and declined in levelized costs by 70\% and 90\% over the past ten years, respectively \citep{Creutzig2017,Weibezahn2022}. In some countries, wind and PV already constitute a major share of power generation, for instance, 50\% in Denmark or 32\% in Germany in 2021 \citep{owidenergyb}. However, further increasing these shares and completely phasing out fossil fuels, planned in Germany until 2035, remains a challenge, because wind and PV power are weather dependent and fluctuate over time and location \citep{reuters2022c}. As a result, higher shares require complementary technologies that can flexibly secure supply, like storage systems, carbon-neutral thermal plants, or transmission infrastructure \citep{Schill2020a,Schaber2012}.

Options to utilize electricity as a primary energy source either consume electricity directly or indirectly via synthetic fuels produced from electricity. The specific strategies vary by sector: In residential heating, most policies encourage a shift to electric heat pumps; in transport, to battery electric vehicles (BEVs) and to some extent power fuels \citep{wh2022,ec2022,bbc2022a,reuters2022d}. In the industry, the strategy is to either electrify suitable processes directly or utilize synthetic fuels, most prominently hydrogen \citep{Sorknaes2022}. Energy policy promotes electrification, because it is already cost competitive in a lot of cases, for instance in heating or transport, and non-emitting alternatives are scares \citep{Luderer2021}. Biomass only has a limited sustainable potential between 100 and 300\,EJ, far from sufficient to satisfy global demand \citep{Creutzig2017}. Carbon capturing, if available, will most likely be limited to industrial applications.

Overall, electrification decisively shapes the demand renewable wind and solar must supply. It impacts the total level, the pattern, and the elasticity of electricity demand---and therefore, to what extent flexible technologies, for instance storage, must complement renewables \citep{Heggarty2019}. As a result, different options for electrification set different requirements for electricity supply. Heat pumps for example have a temperature-dependent efficiency that drives up consumption in winter when PV generation is lowest \citep{Waite2020}. Therefore, they require seasonal storage or more investment in wind generation that peaks in winter as well \citep{Ruhnau2020}. Heating with synthetic fuels on the other hand mitigates flexibility needs, but its low efficiency increases the total electricity demand.

In this paper, we analyze how electrification can support the integration of fluctuating renewables and mitigate the need for storage and thermal backup plants. For this purpose, we apply a comprehensive yet highly detailed energy system planning model. It applies an hourly temporal and sub-national spatial resolution to accurately capture fluctuations of renewables. To investigate how demand from direct and indirect electrification can adapt to these fluctuations, capacity in the heat, transport, and industry sector is endogenous to the model and operational restrictions in these sectors are represented in detail. As a result, the model determines a cost-efficient equilibrium between supply- and demand-side options for flexibility. For example, load peaks of electric heat-pumps can either be covered by grid batteries and thermal plants or mitigated by pairing heat-pumps with local heat storage or even switching to other heating systems, like hydrogen boilers or district heating.

Thanks to this methodology, the consideration of flexibility options in our study is comprehensive. Most previous research excludes relevant options creating a positive bias toward the considered alternatives. For instance, several studies on renewable integration are limited to supply-side options and exclude transmission or electrification, potentially overestimating the need for storage and firm capacity \citep{Sepulveda2018,Dowling2020,Ziegler2019,Sepulveda2021}. More advanced analyses do consider electrification---but only in a single sector. For instance, some studies exclusively consider synergies between renewable electricity and the generation of synthetic fuels, most importantly hydrogen \citep{Caglayan2021,Sasanpour2021,Ruggles2021}. Other studies analyze how electric mobility benefits renewable integration \citep{Verzijlbergh2014, Wei2021, Gunkel2020}, or solely focus on flexibility from residential and district heating \citep{Bloess2019, Bergaentzle2019, Bernath2019,Schill2020}. Finally, there are studies that do cover several sectors but limit the analysis to a single region not considering transmission infrastructure \citep{Schill2020b,Jing2022,Bellocchi2020}. 

Compared to the few studies that do consider storage, electrification, and transmission, our analysis is more detailed with regard to the integration of renewables \citep{Brown2018,Pickering2022}. It applies greater spatio-temporal detail to capture fluctuations and novel methods to represent operational restrictions on the consumer level. On this basis, we are able to visualize and quantify the contribution each sector makes to the system's flexibility. 

The remainder of this paper is structured as follows: The next section~\ref{2} describes the general methodology and introduces several innovations to represent operational restrictions and flexibility related to electrification. Afterwards, section~\ref{3} presents the specific case study, a fully renewable European energy system, that the model is applied to. Section~\ref{4} presents the results of the model including an in-depth analysis of system flexibility based on residual load curves. Section~\ref{5} concludes by discussing policy implications and giving an outlook on future work. Finally, the appendix provides additional details on the deployed model and its results.

\section{Methodology} \label{2}

For the analysis, we apply a linear optimization model that decides on the expansion and operation of technologies to satisfy final energy demand. The model's objective is to minimize total system costs consisting of annualized expansion and operational costs for technologies and costs of energy imports from outside the system. Expansion and operation in the model cover both technologies for the generation, conversion, or storage of energy carriers and grid infrastructure to exchange energy between different regions. 

The model deploys a graph-based formulation specifically developed to model high shares of fluctuating renewables and sector integration, which is capable to vary temporal and spatial resolution within a model \citep{Goeke2020b}. Thanks to this feature, high resolutions can be applied where the system is sensitive to small imbalances of supply and demand---for instance, in the power sector, while more inert parts, like transmission of gas or hydrogen, are modeled at a coarser resolution. This method does not only reduce computational complexity but can also capture inherent flexibility in the energy system, for instance in the gas grid.

Eqs.~\ref{eq:1} to \ref{eq:3c} provide a stylized formulation of the underlying optimization problem. In all equations, variables are written in upper- and parameters in lower-case letters. Regarding expansion, the model decides on capacities $K$ for generation, storage, and exchange; with regard to operation on quantities for generation $G$, use $U$, storage $S$, and exchange $E$. The problem's objective in Eq.~\ref{eq:1} minimizes the sum of fixed costs depending on capacities, and variable costs depending on generation. Specific fixed costs $v^{fix}$ include annualized investment costs plus fixed operational costs. To compute total costs, components are summed over all time-steps $T$, regions $R$, technologies $I$, and carriers $C$. 
\begin{equation}
\min_{K, \,  G, \, U, \, S, \, E} \; \sum_{r \in R, i \in I} K_{r,i}^{gen/st/lvl} \cdot v_{r,i}^{fix} + \sum_{r \in R, r' \in R, c \in C} K^{exc}_{r,r',c} \cdot v_{r,r',c}^{fix}  + \sum_{t \in T, r \in R, i \in I, c \in C} G_{t,r,i,c} \cdot  v_{t,r,i,c}^{var} \label{eq:1}
\end{equation}

Eqs.~\ref{eq:2a} to \ref{eq:2d} list the capacity restrictions that constrain the operational variables by connecting them to capacities. Eq.~\ref{eq:2a} limits the generation $G$ to the installed capacity $K^{gen}$ corrected with the capacity factor $\alpha$ that reflects the share of capacity currently available. Analogously, Eq.~\ref{eq:2b} restricts the storage in- and outflow, $S^{in}$ and $S^{out}$, to the storage power capacity $K^{st}$; Eq.~\ref{eq:2c} restricts the storage level $S^{lvl}$ to the energy capacity $K^{lvl}$. In Eq.~\ref{eq:2d} the capacity of the transmission infrastructure $K^{exc}$ limits the exchange of energy $E$ where the first subscript $r$ refers to the exporting and the second subscript $r'$ to the importing region.
\begin{subequations} 
\begin{alignat}{3}
\sum_{c \in C} G_{t,r,i,c} & \; \; \leq \; \; & \alpha_{t,r,i} \cdot K^{gen}_{r,i}  \hspace{0.35em} & \; \;  \forall t \in T, r\in R, i \in I_{te} \label{eq:2a} \\  
\sum_{c \in C} S^{in}_{t,r,i,c} + S^{out}_{t,r,i,c}  & \; \; \leq \; \; & K^{st}_{r,i} \hspace{0.85em} & \; \; \forall t \in T, r\in R, i \in I_{st}   \label{eq:2b} \\ 
 S^{lvl}_{t,r,i} & \; \; \leq \; \; & K^{lvl}_{r,i} \hspace{0.6em} & \; \;  \forall t \in T, r \in R, i \in I_{st}  \label{eq:2c} \\
E_{t,r,r',c} + E_{t,r',r,c}  & \; \; \leq \; \; & K^{exc}_{r,r',c} & \; \;  \forall t \in T, r \in R, r' \in R, c \in C \label{eq:2d}
\end{alignat}
\end{subequations}

Finally, the balances in Eqs.~\ref{eq:3a} to \ref{eq:3c} restrict the operational variables. First, the energy balance in Eq.~\ref{eq:3a} ensures that supply meets the demand $d$ at all times, in each region, and for each energy carrier. Eq.~\ref{eq:3b} controls how technologies convert energy carriers setting the amount of generated energy $G$ to the product of utilized energy $U$ and the efficiency $\eta$. The storage balance in Eq.~\ref{eq:3c} tracks the storage level $S^{lvl}$ which connects the storage balance in the previous period $t-1$ plus in- and minus outflows. The parameters $\delta$ and $\rho$ reflect self-discharge and charging losses, respectively.
\begin{subequations}
\begin{alignat}{3}
 \sum_{i \in I} (G_{t,r,i,c} - U_{t,r,i,c} + S^{out}_{t,r,i,c} - S^{in}_{t,r,i,c})  + \sum_{r \in R'} (E_{t,r,r',c} - E_{t,r',r,c}) & \; \;  = \; \; & d_{t,r,c}  & \;  \; \forall t \in T, r \in R, c \in C   \label{eq:3a} \\ 
\sum_{c \in C} \eta_{t,r,i} \cdot U_{t,r,i,c} & \; \;  = \; \; & \sum_{c \in C} G_{t,r,i,c}  & \;  \; \forall t \in T, r \in R, i \in I   \label{eq:3b} \\ 
\delta_{t,r,i} \cdot S^{lvl}_{t-1,r,i}  + \sum_{c \in C}  \rho_{t,r,i} \cdot S^{in}_{t,r,i,c} - S^{out}_{t,r,i,c} & \; \;  = \; \;  & S^{lvl}_{t,r,i} & \;  \; \forall t \in T, r\in R, i \in I_{st}  \label{eq:3c} 
\end{alignat}
\end{subequations}

To achieve high detail and comprehensive scope, the model makes several simplifying assumptions that are common in the literature and found not to impose a significant bias on results. First, the model does not consider operational restrictions of individual power plants, like ramping rates or start-up times, and the need for ancillary services, like balancing reserves. Previous studies agree that such operational detail has little impact on results if models include options for short-term flexibility, like batteries or demand-side response, and conclude modeling of renewable systems should rather prioritize temporal and spatial detail
\citep{Poncelet2016,Poncelet2020,Priesmann2019, Helisto2021}. Second, the model uses a transport instead of a power flow formulation to represent grid operation. Previous research found this simplification to be sufficiently accurate \citep{Neumann2020a}. In addition, model parameterization uses 
net-transfer capacities (NTCs) already reflecting power flow restrictions instead of physical grid capacities, as detailed in section~\ref{3}.

The following two subsections describe how the model captures operational restrictions and flexibility related to electrification in heating and transport extending the stylized model formulation above. Model specifics beyond the mathematical formulation, like considered regions, technologies, and sectors, will follow in section~\ref{3}. The appendix and the linked supplementary material provide more detailed information and the code of the underlying open-source modeling framework AnyMOD \citep{Goeke2020a}.

\subsection{Operation of heating systems} \label{heatSys}

The different technological options to electrify the supply for process and space heat also affect the flexibility of electricity demand differently. Indirect electrification using synthetic fuels is generally the most flexible. Electrolysis or other processes can easily adapt to renewable supply and store their products for later consumption. In contrast, the direct use of electricity with heat-pumps or electric boilers is more energy-efficient but also more constrained. Specific constraints differ depending on whether systems provide industry, residential, or district heat and operate in combination with heat storage. In the following, we describe how the model captures these different constraints.

First, applying the graph-based approach, industry, residential, and district heat use a four-hour resolution representing their inherent flexibility. This means, that while the energy balance is an hourly constraint for electricity, for heat, supply does not have to equal demand in each hour but over the sum of four hours. In residential heating, this resolution captures the thermal inertia of buildings \citep{Heinen2017}; in district heating, the inertia of the network itself \citep{Triebs2022}; and in industrial heating, the possibility to reschedule processes.

\begin{figure}[!htbp]
	\centering
		\includegraphics[scale=0.13]{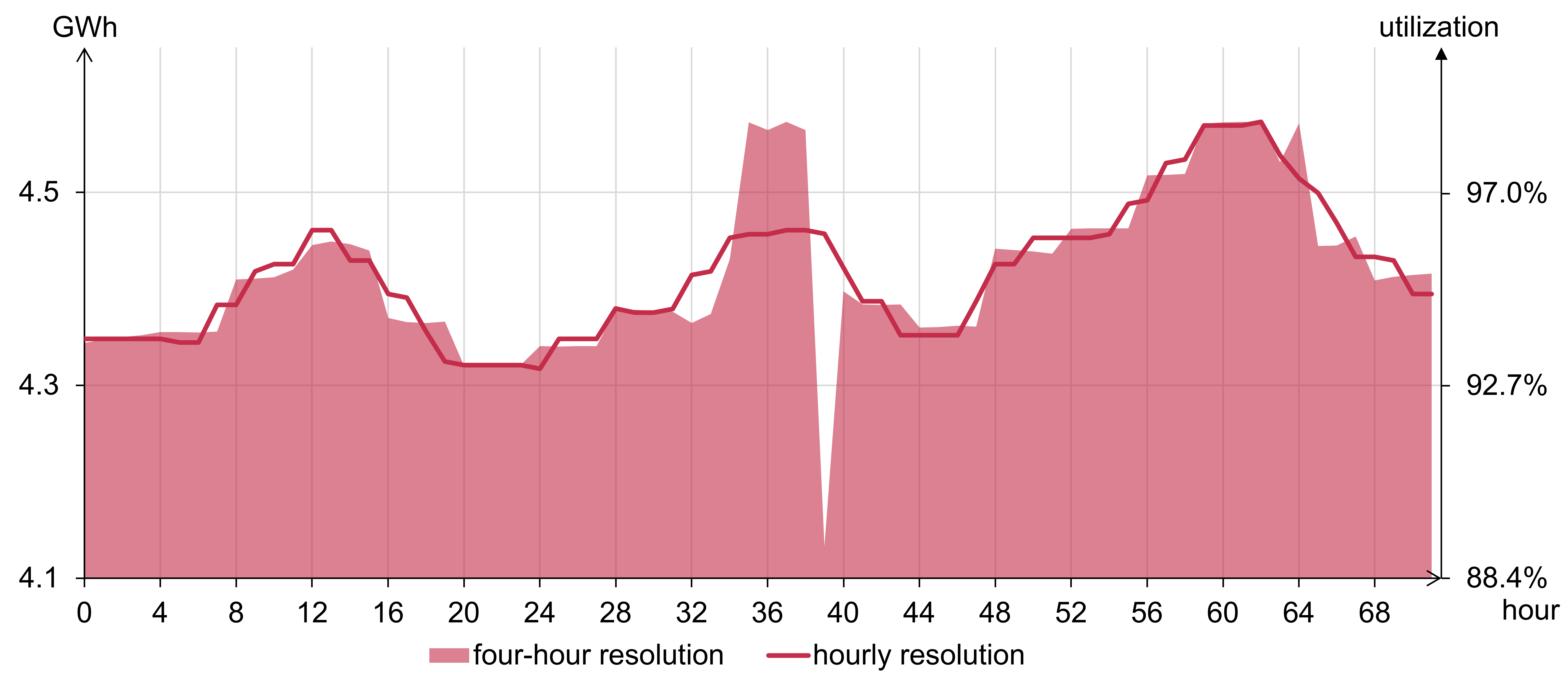}
	\caption{Exemplary electricity demand resulting from different resolutions for process heat}
	\label{fig:deployHeat}
\end{figure}

Fig. \ref{fig:deployHeat} illustrates the concept comparing the electricity demand of an electric boiler providing process heat for an hourly and four-hour resolution. In the hourly case, the electricity demand of the boiler is fixed according to the hourly time-series. In the four-hour case, the demand is flexible but still subject to two constraints. The total demand for each four-hour period must equal the sum of the hourly demand. Consequently, in Fig. \ref{fig:deployHeat} areas above and below the hourly time-series are equal in size in each four-hour period. And, even with a four-hour resolution for heat, electricity demand is still an hourly variable and subject to an hourly capacity constraint. This constraint's impact greatly depends on the total utilization within the specific time-step. For instance, in the four-hour period from hours 36 to 40, the electric boiler must achieve an average utilization rate of 97.5\%. As a result, the most flexible operation possible is to operate at 90\% utilization in one hour and at full capacity in the others.

Second, the model introduces a different operational concept for industrial and residential heating systems. In contrast to the power sector, systems for industrial or residential heating are locally bound and do not feed generation into a transmission network. Instead, they must directly match local demand. On this small-scale, the operation of base- and peak-load plants is not cost efficient, and so the profile of demand directly dictates operation.

To account for this restriction, the model has two different ways to describe the operation of technologies. Eqs.~\ref{eq:4a} and \ref{eq:4b} provide the formulation for the common case of technologies interacting within a network. The formulation consists of an energy balance that ensures the summed generation $G$ from all technologies $i$ equals the demand $d$ at each time-step $t$ and a capacity constraint that limits the output of each technology to the installed capacity $K$ in each time-step $t$. In the problem formulation above, these constraints correspond to Eq.~\ref{eq:3a} and Eq.~\ref{eq:2a}, respectively.
\begin{subequations}
\begin{alignat}{4} 
\sum_{i \in I} \mathit{G}_{i,t} = \; & \; \mathit{d}_{t} &  \hspace{0.85em}  \; \forall t \in T \hspace{2.65em} \label{eq:4a} \\
\mathit{G}_{i,t} \leq \; & \; \mathit{K}_{i} &   \hspace{0.85em} \; \forall t \in T, \, i \in I \label{eq:4b}
\end{alignat}
\end{subequations}

In contrast, Eqs.~\ref{eq:5a} and \ref{eq:5b} describe the case of unconnected technologies each operating to match local demand individually. Instead of an energy balance for each time-step, a single capacity balance ensures that the installed capacity can meet peak demand $p$. The second equation fixes the operation of technologies in each time-step according to the demand profile, which corresponds to the ratio of current demand to peak demand. Since this formulation replaces the energy balances for each time-step with a single constraint on capacity and the inequality constraint for capacity with an equality constraint, it also reduces the complexity of the optimization model.
\begin{subequations}
\begin{alignat}{3}
\sum_{i \in I} \mathit{K}_{i} = \; & \; \mathit{p} & \label{eq:5a} \\
\mathit{G}_{i,t} = \; & \; \frac{\mathit{d}_{t}}{p} \cdot \mathit{K}_{i}  &  \hspace{0.85em}  \; \forall t \in T, \, i \in I \label{eq:5b}
\end{alignat}
\end{subequations}

\begin{figure}[!htbp]
	\centering
		\includegraphics[scale=0.13]{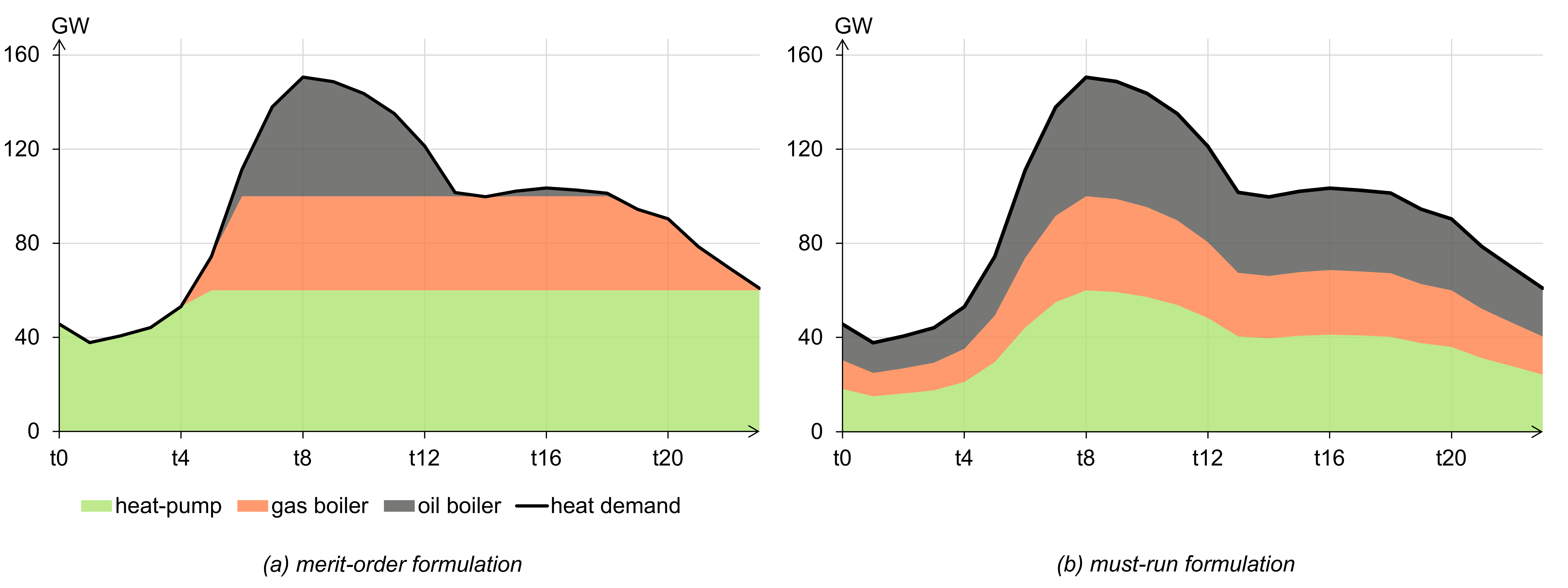}
	\caption{Illustration of deployment concepts}
	\label{fig:deployment}
\end{figure}

Fig.~\ref{fig:deployment} shows the operation of the same capacities for each formulation to illustrate their differences. In the first case, a cost minimizing model will automatically operate capacities according to the merit order, hence the name merit-order formulation. With increasing demand, technologies are successively deployed in the order of their marginal costs. In the second formulation, termed must-run, all technologies must run simultaneously with generation shares corresponding to capacities. 
In industrial or residential heating, at a given capacity the merit-order formulation will overestimate the generation share of technologies with low marginal costs and vice versa underestimate the generation of technologies with high marginal costs. As a result, implausible investment into "peak-load" capacities can occur. For instance, residential hydrogen boilers could be built to run at very low utilization and provide additional heat when electricity for heat pumps is scarce. However, in practice, this is implausible implying consumers install two redundant heating systems in their homes.

Accordingly, in the model, technologies providing process or space heat use the must-run formulation; technologies providing electricity, hydrogen, and synthetic gas the merit-order formulation. District heat uses the merit-order formulation as well to capture how the operation of different plants within heating networks is flexible. On the other hand, substations that transfer district heat to final industrial or residential consumers and determine the demand for district heating use the must-run formulation. 

Finally, the model can invest in heat storage to add flexibility. For district heating, the implementation of storage is straightforward and analogous to carriers like electricity or hydrogen. For space and process heating, viz. local technologies using the must-run formulation, the storage is directly embedded into the specific heating technology. This setup again prevents inconsistencies resulting from the interplay of unconnected local technologies, like heat storage charged by hydrogen boilers but discharging to consumers with heat-pumps.

\begin{figure}[!htbp]
	\centering
		\includegraphics[scale=0.5]{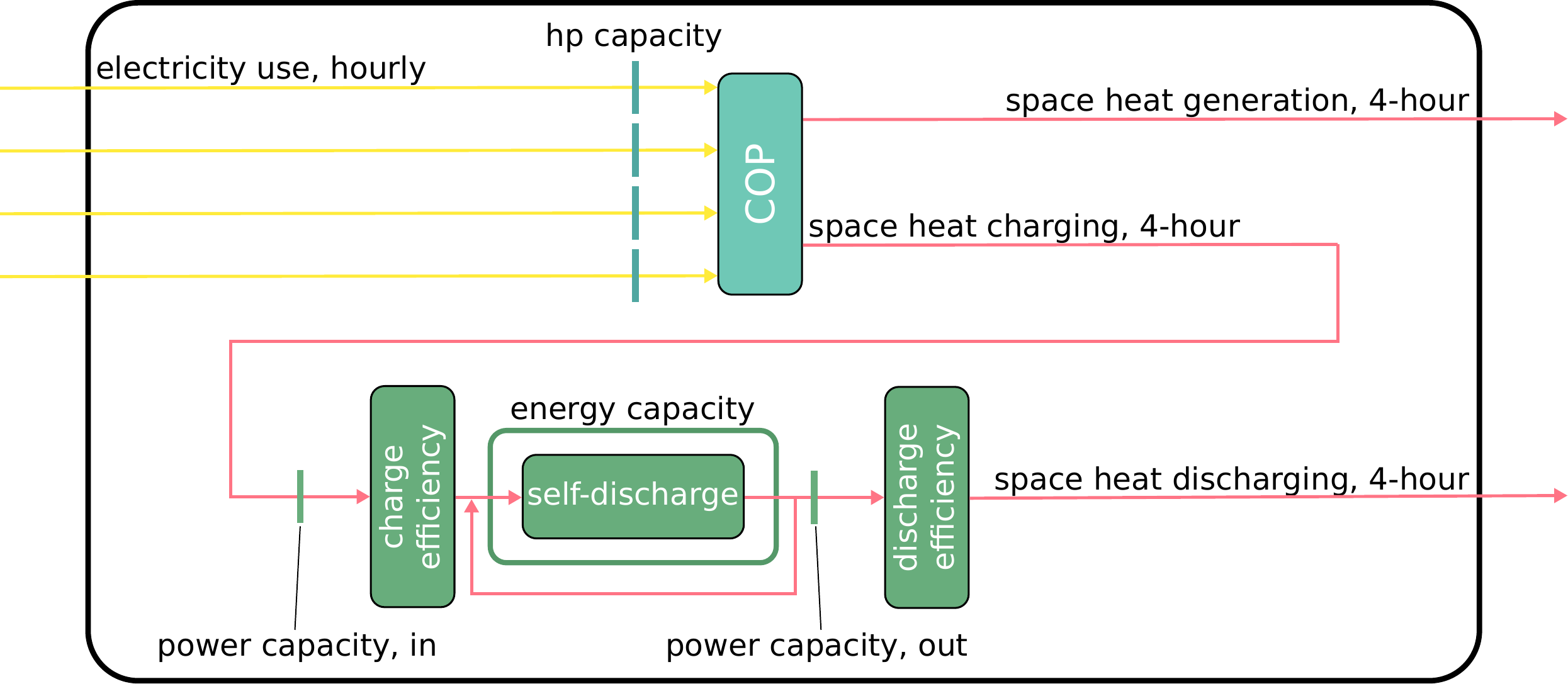}
	\caption{Representation of residential heat-pump paired with local heat storage}
	\label{fig:hp}
\end{figure}

Fig.~\ref{fig:hp} describes this concept for an electric heat-pump. The upper row illustrates how the heat-pump converts electricity to heat in a ratio equal to the coefficient of performance (COP). The vertical lines indicate the capacity constraints imposed on the hourly electricity demand. The generated heat, modeled at a four-hour resolution, can either cover demand or be transferred to the storage system displayed in the lower part of the figure. The level of storage investment determines the power and energy capacity of the storage. Storage losses depend on the self-discharge rate, the charge, and the discharge efficiency.  
Only heat leaving the system boundary, indicated by the black line in Fig.~\ref{fig:hp}, adds to the must-run output referenced in Eq.~\ref{eq:2b}. Discharging the storage enables the heat-pump to produce less than the current must-run output but requires previous charging. In this way, storage can reduce demand when electricity is scarce at the cost of storage losses increasing overall demand. Due to high discharge rates, long storage durations are not viable with local heat storage.

\subsection{Charging of BEVs}

The flexibility of BEVs in future energy systems is still subject to uncertainty and depends on technological and regulatory developments. In this study, we make a middle-ground assumption on the flexibility of electricity demand from BEVs. On the one hand, we assume charging to be flexible within limits and can adapt to supply which does not reflect current regulation in all European countries but neither requires additional infrastructure \citep{Strobel2022}. On the other hand, we do not assume that BEVs can feed electricity back to the grid, also termed bidirectional charging or vehicle-to-grid, which requires bidirectional chargers \citep{Hannan2022}.

\begin{figure}[!htbp]
	\centering
		\includegraphics[scale=0.13]{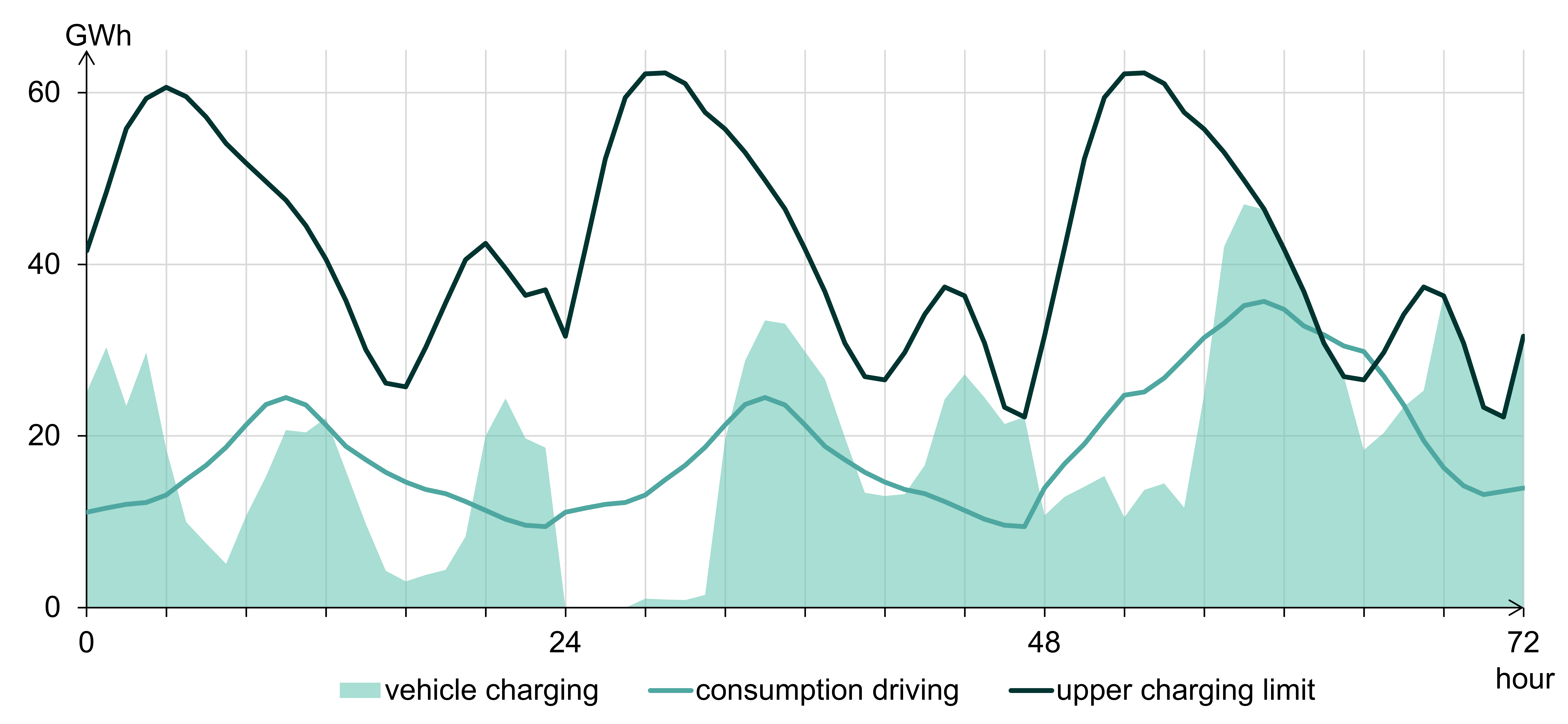}
	\caption{Exemplary electricity demand of BEVs}
	\label{fig:deploymentBEV}
\end{figure}

Fig.~\ref{fig:deploymentBEV} demonstrates how the model implements flexible charging based on an exemplary driving and charging pattern for private passenger cars. First, an hourly profile restricts the charging of BEVs to reflect the capacity of vehicles currently connected to the grid. Second, on each day the electricity charged must match the consumption for driving. Consequentially, areas above and below the curve for consumption are equal in size for each day, analogously to Fig.~\ref{fig:deployHeat}. Instead of explicitly modeling battery levels, this approach implicitly assumes that on average batteries can balance charging and consumption over one day at least. 

To implement the approach, transport services use a daily resolution in the model. Conversely, other electricity demand for transport, for instance for rail transport, is inflexible and uses an hourly resolution.

\section{Case study} \label{3}

\begin{figure}[!htbp]
	\centering
		\includegraphics[scale=0.85]{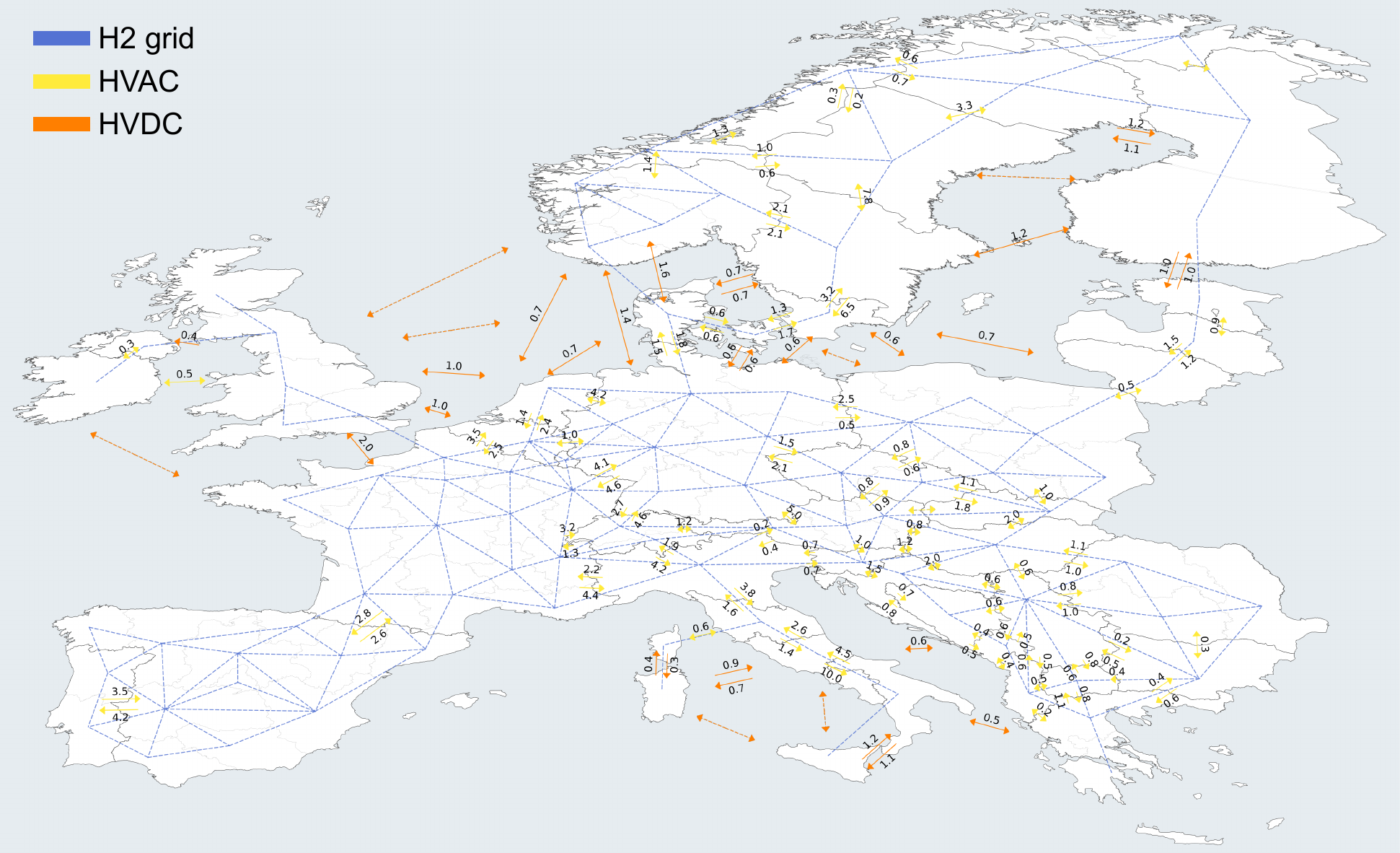}
	\caption{Pre-existing grid infrastructure (solid) and new potential connections (dashed) in the model}
	\label{fig:startGrid}
\end{figure}

To study the impact of electrification on renewable integration, we apply the outlined planning model to a fully renewable European energy system. Thanks to this large spatial scope, mapped in Fig.~\ref{fig:startGrid}, the analysis accounts for transmission infrastructure and the exchange of energy as one option to integrate renewables. A spatial resolution with 96 clusters defined by the dotted lines captures local fluctuations of renewable generation. 

Blue lines indicate where the model can invest in grid infrastructure for hydrogen between clusters at costs of 0.4\,Mil. € per GW and km and energy losses of 2.44\% per 1,000\,km \citep{DEA}. The distance between the geographic center of clusters serves as an estimate for pipeline length.

\begin{figure}[!htbp]
	\centering
		\includegraphics[scale=0.13]{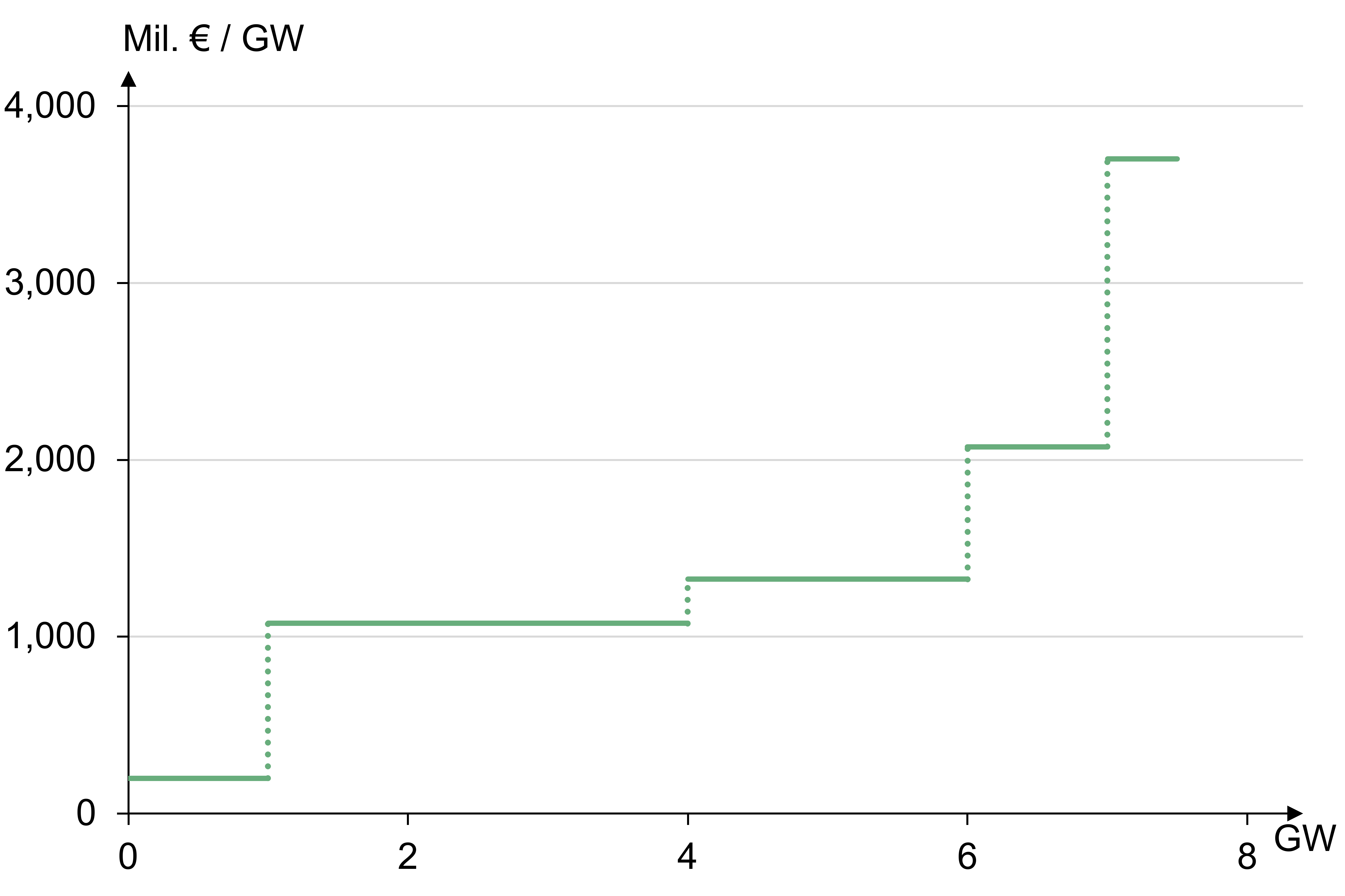}
	\caption{Exemplary capacity-cost curve for NTCs between Germany and the Netherlands}
	\label{fig:ntcExp}
\end{figure}

The representation of the power grid aggregates clusters according to the zones of the European power market. Yellow and orange arrows indicate pre-existing high-voltage alternating current (HVAC) and direct current (HVDC) connections. Arrows without a number indicate a potential connection without pre-existing capacity. Building on data in \citet{entsoe2020}, the expansion of specific connections is subject to a capacity-cost curve. Fig.~\ref{fig:ntcExp} exemplarily shows this curve for the NTC between Germany and the Netherlands. In this case, the specific investment costs of the NTC discretely increase from 200 to 3,700 Mil. € per GW and expansion is subject to an upper limit of 7.5\,GW. Transmission losses amount to 5\% and 3\% per 1,000\,km for HVAC and HVDC, respectively \citep{Neumann2020a}. 

The temporal scope of the model consists of a single year. Using a brownfield approach, today's transmission infrastructure and hydro power plants are available without expansion. In total, the applied model includes 22 distinct energy carriers that can be stored and converted into one another by 120 different technologies covering heating, transport, industry, and the production of synthetic fuels. Section~\ref{a} of the appendix provides comprehensive documentation.

Fig.~\ref{fig:powerAll} provides an overview of technologies for electricity generation. Vertices in the graph either represent energy carriers, depicted as colored squares, or technologies, depicted as gray circles. Entering edges of technologies refer to input carriers; outgoing edges refer to outputs. Extraction turbines and biomass plants can be operated flexibly decreasing their heat-to-power ratio at the cost of reduced total efficiency. Both reservoirs and pumped storage operate as storage, but reservoirs are charged based on an exogenous time series while charging of pumped storage is endogenous.

\begin{figure}[!htbp]
	\centering
		\includegraphics[scale=0.33]{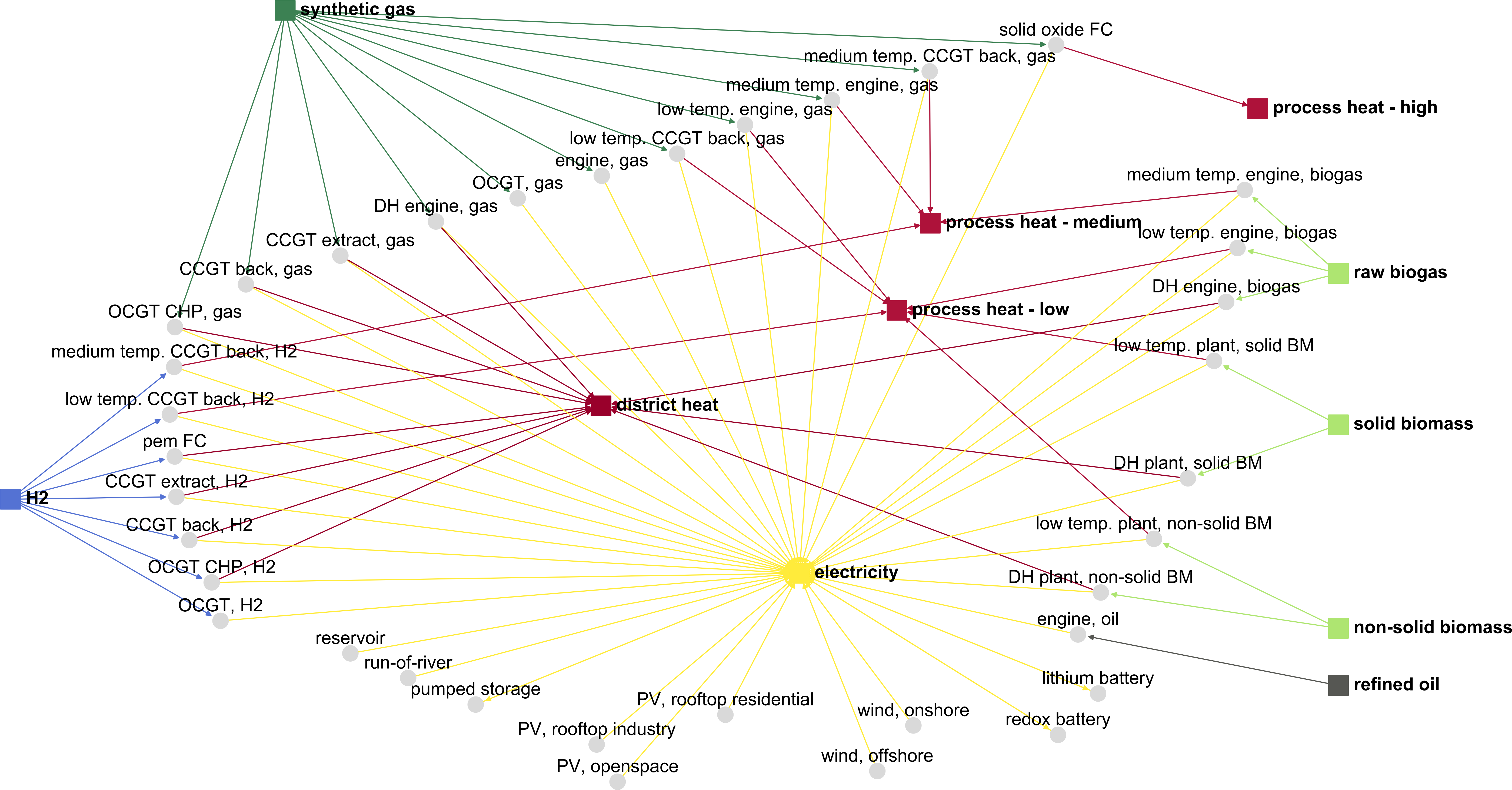}
	\caption{Subgraph for electricity supply}
	\label{fig:powerAll}
\end{figure}

The choice of technologies and their parameterization are based on the reports by the \citet{DEA}, except for transport where data comes from \citet{Robinius2020}. For hydrogen fueled power plants, we assume a 15\% mark-up on the costs of the corresponding natural gas technology in line with \citet{Oberg2022}. Section~\ref{b} of the appendix lists all technology data for the power sector.

The capacity and energy potential of PV and wind are differentiated according to the 96 clusters displayed in Fig.~\ref{fig:startGrid}. In addition, openspace PV and onshore wind are further broken down into three categories with different full load hours for each cluster to reflect different site qualities; rooftop PV and offshore wind are broken into two further categories. Capacity limits are scaled to comply with the overall energy potential for each country reported in \citet{Auer2020}. Time-series data for capacity factors is, like all time-series data, based on the climatic year 2008 \citep{Osmose}.
 
BEVs for private passenger and light freight transport have a charging capacity of 5\,kW; BEVs for public passenger and heavy road transport of 150\,kW \citep{entsoe2021}. Applying a safety margin all charging profiles are reduced by 75\%. 

To estimate the technical potential of different space heating technologies, we use national Eurostat data on urbanization \citep{eurostat2022}. For rural areas, we assume that ground- and water-source heat-pumps can cover the entire demand, but district heating is not available. Vice versa, for cities, district heating can cover the entire demand, but heat-pumps are not available. For towns and suburbs, district heating, ground-source, and water-source heat-pumps can each cover 50\% of demand. To estimate the technical potential of different process heating technologies, technology info from \citet{DEA} on eligibility for different processes is paired with national data on industry activity \citep{jrc2012,eurostat2022b}.

The use of biomass in each country is subject to an upper energy limit that sums to 1,081\,TWh for the entire model \citep{jrc2015}. In addition to domestic production, the model can import renewable hydrogen by ship at costs of 111.71\,€ per MWh and by pipeline from Morocco or Egypt at 76.9 and 73.56\,€ per MWh, respectively \citep{Hampp2021}.

\section{Result} \label{4}

To give an impression of the resulting energy system, section~\ref{enBal} summarizes the energy flows and balances when solving the model for the described case study. Since these results are largely in line with previous studies, the main purpose is to provide a context for the subsequent section~\ref{durCur} that closely analyses the system integration of renewables and the contribution of flexible electrification.

\subsection{Energy balances} \label{enBal}

\begin{sidewaysfigure}
	\centering
		\includegraphics[scale=0.5]{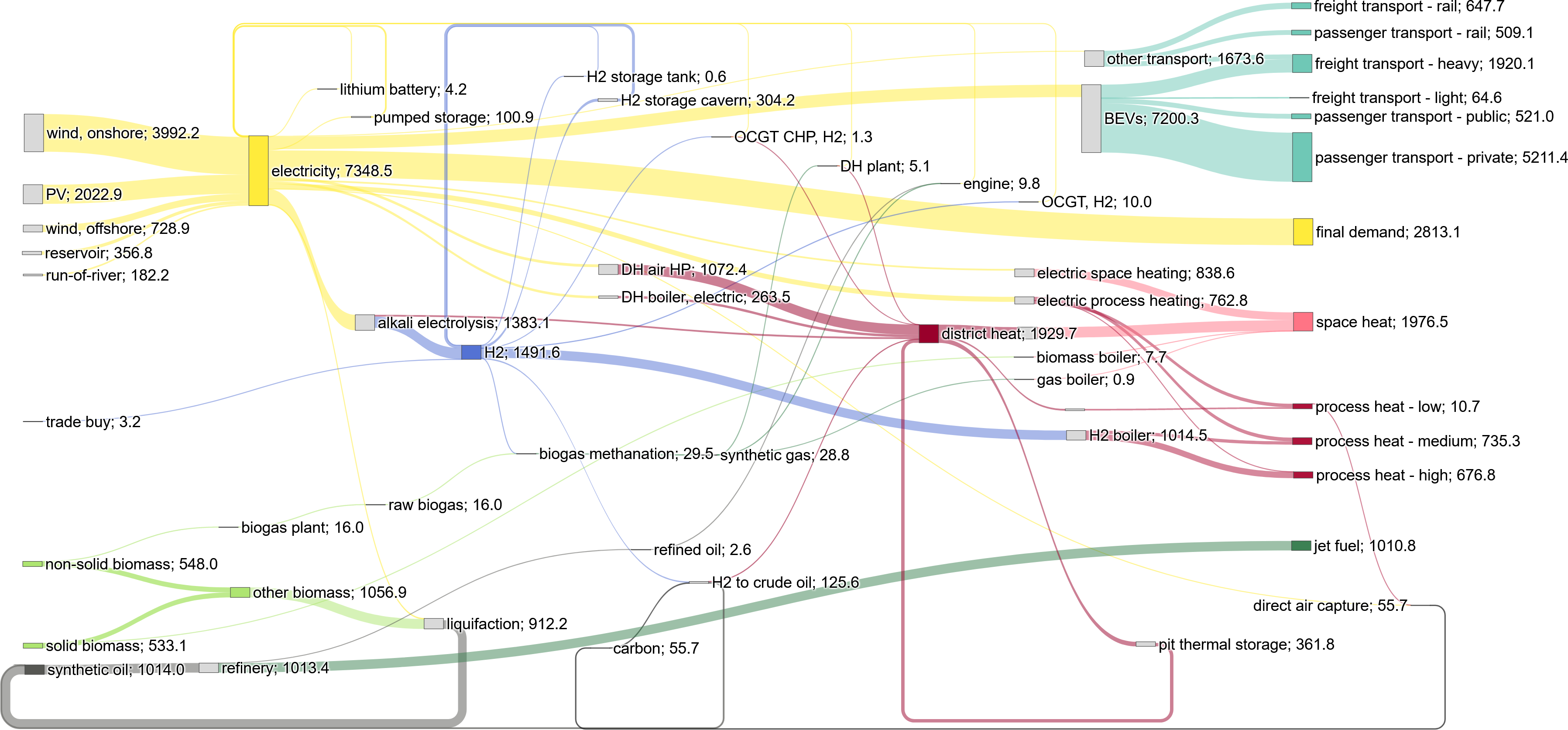}
	\caption{Sankey diagram for the solved model, in TWh/Gpkm/Gtkm}
	\label{fig:sankeyGrid}
\end{sidewaysfigure}

The Sankey diagram in Fig.~\ref{fig:sankeyGrid} illustrates energy flows in the solved model. On the right side, the diagram shows the final demand for energy and transport services. Going from right to left, the diagram details how the model deploys conversion processes, storage, and secondary energy carriers to meet the demand from primary energy sources. To be clear and concise, the diagram aggregates individual technologies, like different types of BEVs, into one node. The ratio of flows entering and leaving a node reflects the average efficiency of the underlying technology. For instance, heat-pumps have efficiencies greater than one and the generated heat exceeds the electricity consumed, so outgoing flows exceed incoming flows. Technologies can have multiple in- and outputs, like alkali electrolysis, which produces hydrogen but also provides waste heat to district heating networks. For storage, incoming and outgoing flows relate to the same energy carrier, though outgoing flows are smaller reflecting storage losses.
Overall, the results show a clear emphasis on direct electrification in the heating, transport, and industry sector. In space and district heating, heat-pumps and electric boilers provide 99.5\% of the total demand. The transport sector uses BEVs and overhead lines wherever possible and synthetic carriers only cover the exogenous demand for transport fuels. Although indirect electrification using hydrogen is an option to create these fuels, the model predominantly utilizes the available biomass potential instead. Overall, indirect electrification is only relevant in process heating above 100\degree{}C, which accounts for 85\% of the total hydrogen demand. This is partly due to the limited potential of direct electrification at these temperature levels. Nevertheless, only 73\% of the electrification potential is utilized suggesting that in some cases the model deploys indirect over direct electrification despite its inefficiency because it is more flexible. Generating and storing hydrogen for later use is comparatively easy but there are no options for heat storage above 100\degree{}C and the operational flexibility in process heating is small. 

While the general prevalence of direct electrification aligns with previous research, our results deviate in several details. First, the endogenous share of district heating is close to the upper limit and exceeds fixed shares in previous studies \citep{Bloess2019,Brown2018}. Our methodology capturing how district heating is more flexible than individual heating presumably drives these results. Second, thermal plants only provide 25.6\,TWh of firm generation, much less than in previous deep decarbonization studies limited to the power system and a single region \citep{Sepulveda2018,Ziegler2019}. This suggests electrification and transmission greatly contribute to renewable integration and substitute thermal backup plants.

\begin{figure}[!htbp]
	\centering
		\includegraphics[scale=0.85]{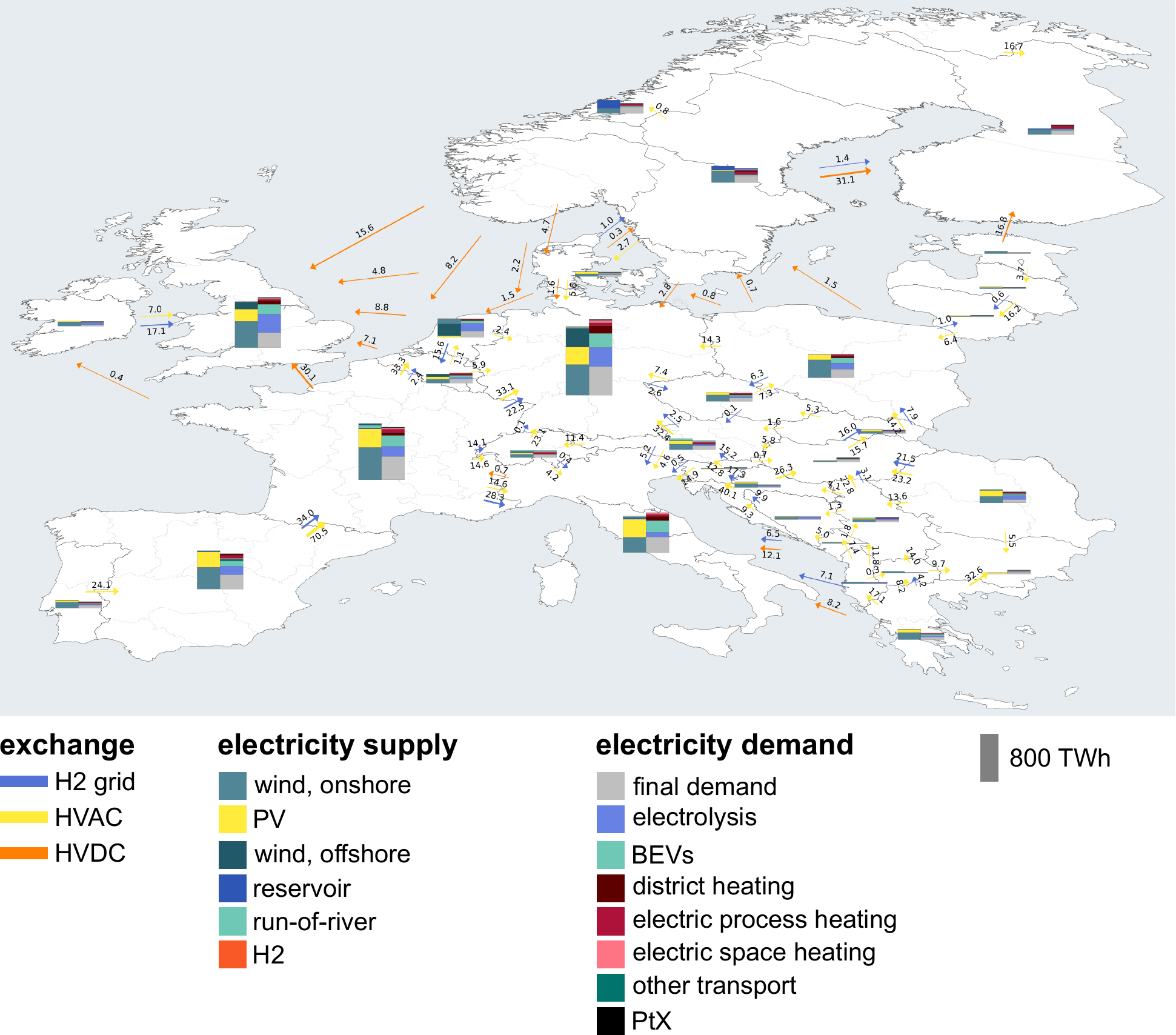}
	\caption{Net-exchange (in TWh), electricity generation and demand}
	\label{fig:resMap}
\end{figure}

The results mapped in Fig.\,\ref{fig:resMap} highlight the importance of transmission. The figure shows the net-exchange for different transmission infrastructures and electricity generation and demand by country.

In the electricity grid, 70\% of the potential grid expansion is realized. NTC capacities for HVAC more than double from 109 to 275\,GW and quadruple for HVDC from 19 to 75\,GW. Traded quantities increase correspondingly, but net positions remain comparatively balanced. For instance, Germany has net-imports of 124.5\,TWh in the results compared to net-exports of 17.4\,TWh in 2021, but nevertheless exports double compared to 2020 and amount to 115.2\,TWh \citep{smard}. This indicates that a key driver of grid expansion is to balance local fluctuations of renewable supply.

The results for the exchange of hydrogen are opposed and trade is much more unilateral. Some countries, like Spain or Romania, have a comparative advantage in producing hydrogen due to high capacity factors and are exclusively exporting. Other countries, like Italy, serve as intermediaries or are exclusively importing, like Belgium. Overall, domestic hydrogen production is cost-efficient and hydrogen imports from outside of Europe are negligible totaling 3.2\,TWh imported by Italy and the UK.

\subsection{Residual load curves} \label{durCur}

While the previous section only indicates how the system achieves the integration of fluctuating renewables, this section provides a definite analysis based on residual load curves.

Residual load curves show total demand minus fluctuating renewable generation sorted in descending order and depict the energy that sources other than fluctuating renewables must supply \citep{Schill2014}. The y-axis intercept of the curve gives the residual peak load, the highest capacity from other sources. The area above the x-axis corresponds to the amount of energy from other sources; the area below the x-axis to excess generation.

\begin{figure}[!htbp]
	\centering
		\includegraphics[scale=0.13]{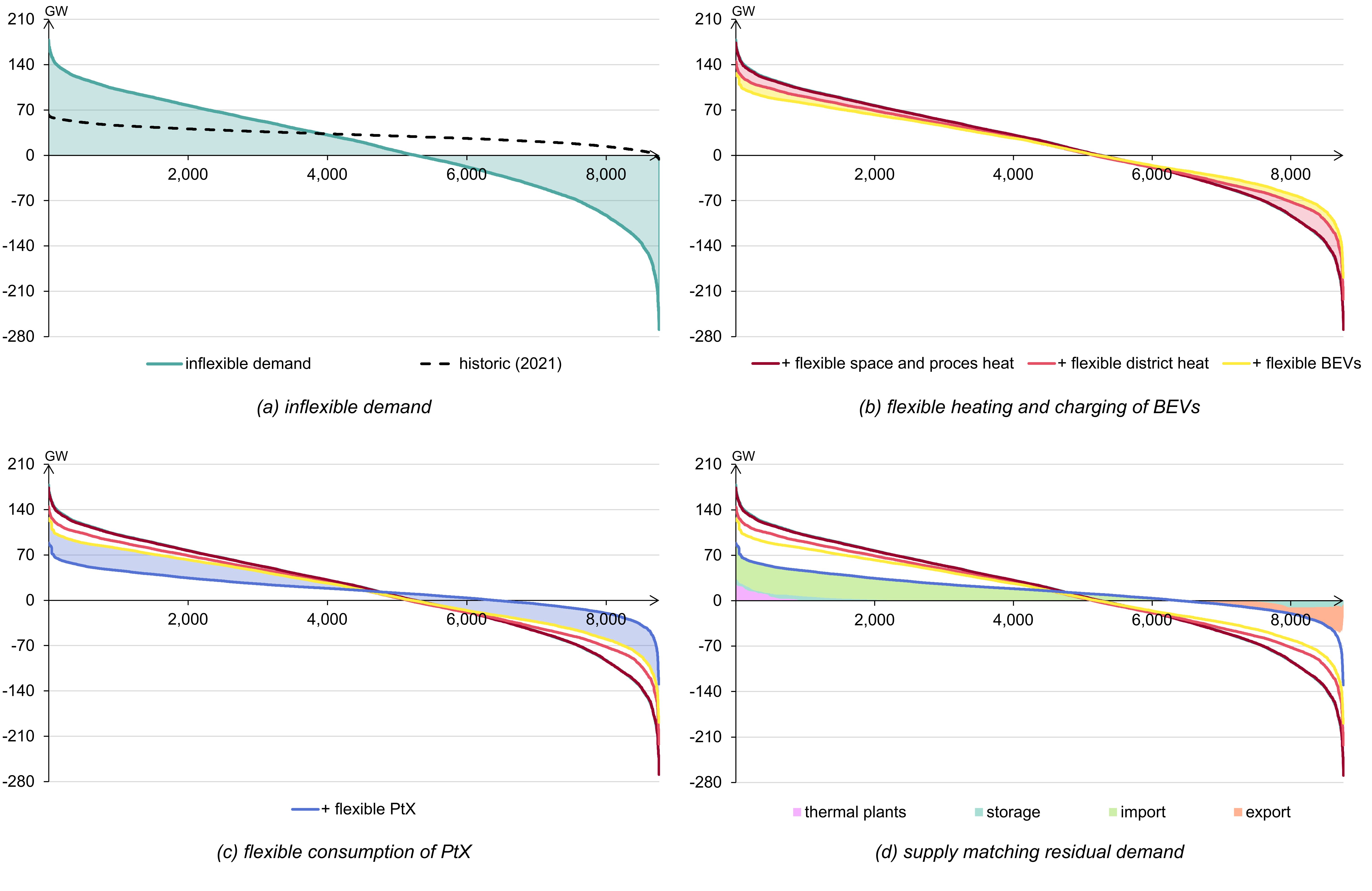}
	\caption{Residual load curves for Germany \citep{energyCharts}}
	\label{fig:dcGridDe}
\end{figure}

Fig.~\ref{fig:dcGridDe}(a) compares residual load in 2021 and in the model results if demand were completely inflexible. The inflexible demand is not a direct result of the model but computed ex-post assuming electric heating without the flexibilities described in section~\ref{heatSys}, charging of BEVs proportional to loading profiles, and power-to-x (PtX) processes, mostly electrolyzers, operating at constant capacity. For illustration, Fig.~\ref{fig:ts} in the appendix~\ref{c} provides a section on the time-series data the residual load curves are based on. Exemplary for the entire system, the analysis focuses on Germany, the country with the highest demand and a relatively small renewable potential. 

The comparison of historic values and model results in Fig.~\ref{fig:dcGridDe}(a) shows a dramatic increase in residual load when moving to a renewable energy system without flexible demand. Compared to 2021, residual peak-load almost triples from 62.9 to 178.2\,GW and residual demand amounts to 338.3\,TWh, but fluctuating generation also exceeds demand in 3,491 hours resulting in 205.5\,TWh of excess generation.

Fig.~\ref{fig:dcGridDe}(b) illustrates how flexible heating and charging of BEVs reduce residual demand. To compute these curves, inflexible demand computed ex-post is successively replaced by hourly demand from model results; first for flexible space and process heating, then for district heating, and finally for BEVs. This order is arbitrary and serves the communication of the results.

The impact of flexible space and process heating is small and only reduces residual peak-load by 4.3\,GW and residual demand by 2.0\,TWh. Correspondingly, the model does not invest in local heat storage. The influence of flexible district heating is much more pronounced reducing peak-load by 29.4\,GW and residual demand by 38.7\,TWh. At the same time, excess generation only decreases by 36.3\,TWh because heat-pumps can shift operation to periods with higher efficiencies. To achieve this flexibility, the model builds 40.5\,TWh of thermal water storage and combined heat and power (CHP) gas engines with 3.0\,GW heating capacity. The effect of flexible BEV charging is significant as well. Residual peak-load decreases by 18.2\,GW and residual demand by 28.7\,TWh.

The flexible operation of electrolyzers displayed in Fig.~\ref{fig:dcGridDe}(c) has the greatest impact on residual demand. When electrolyzers adapt to supply instead of operating at constant capacity, residual peak-load decreases by 37.8\,GW and residual demand by 98.1\,TWh. In this case, 82.8\,GW of electrical electrolyzer capacity operate at a utilization rate of 44.4\%. To match volatile hydrogen production with demand, the model invests in 13.0\,TWh of hydrogen storage in salt caverns, still only 0.1\% of the total storage potential in Germany \citep{Caglayan2020}.

In total, flexible electrification has a substantial effect reducing residual peak-load from 178.2 to 88.5\,GW, residual demand from 338.3 to 170.8\,TWh, and excess generation from 205.5 to 41.9\,TWh. Fig.~\ref{fig:dcGridDe}(d) finally shows how the system meets the residual demand. In line with the results in the previous section, transmission covers the major portion. At peak-load Germany imports 52.8\,GW, which corresponds to 96\% of the total import capacity. Imports also cover 88.3\,TWh of the residual demand and exports reduce excess generation by 28.7\,TWh.\footnote{These numbers are smaller than total imports and exports stated in the previous section because they are the sum of net-positions in each hour.} Thermal plants and energy storage are less important. Thermal plants cover 25.2\,GW of peak-load and provide 8.8\,TWh of generation, almost equally divided between open-cycle (OC) hydrogen turbines and gas engines. For electricity storage, the model does not invest in batteries and only deploys the pre-existing hydro plants, which cover 10.5\,GW of peak-load and 8.7\,TWh of residual demand. Charging of hydro plants reduces excess generation by 9.9\,TWh resulting in only 3.3\,TWh of excess generation being finally curtailed.

In the presented scenario, the power grid is the greatest source of flexibility, both on the supply and demand side, and NTC capacities almost triple. But grid expansion frequently faces public opposition and under extreme weather conditions total generation can be insufficient to balance out local shortages. To check the robustness of our results against this background, we solve the model for an additional scenario without any grid expansion. General results without grid expansion do not differ substantially from the reference scenario. Annualized costs of the energy system increase by 5.8\% from 251.9 to 266.6\,bil. € and most notably the exchange and use of hydrogen increase to substitute the power grid. Analogously to section~\ref{enBal}, the appendix~\ref{c} provides a Sankey diagram and map with detailed scenario results. 

\begin{figure}[!htbp]
	\centering
		\includegraphics[scale=0.13]{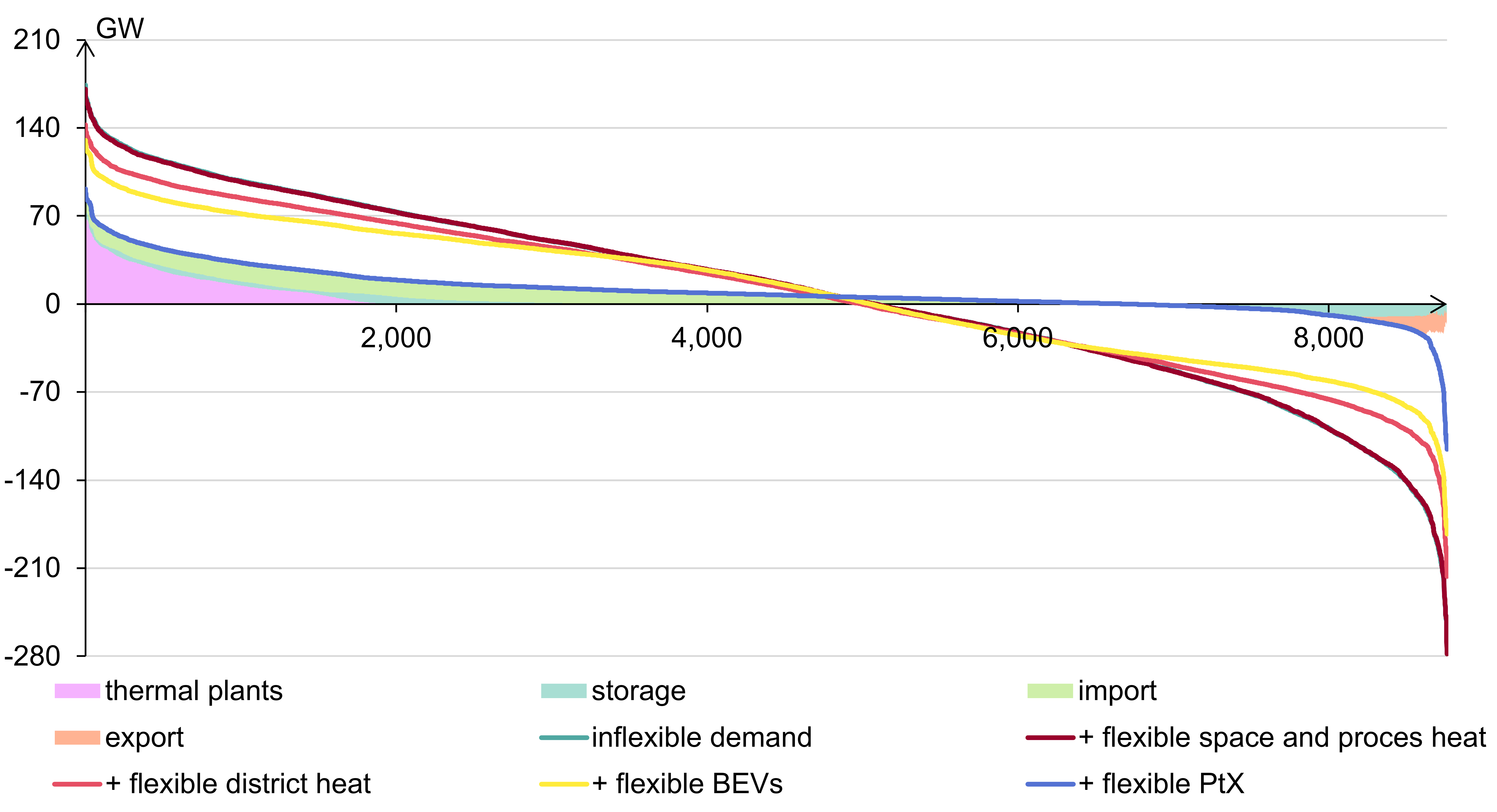}
	\caption{Residual load curves for Germany without grid expansion}
	\label{fig:dcNoGridDe}
\end{figure}

With regard to renewable integration, Fig.~\ref{fig:dcNoGridDe} shows the residual load curves for the scenario without grid expansion. Results on the demand side are similar to the reference scenario and flexible electrification reduces the residual peak-load from 173.9 to 91.4\,GW and the residual demand from 317.7 to 110.0\,TWh. The only substantial difference is a greater contribution from flexible PtX that decreases residual demand by 139.4 instead of 98.1\,TW in the previous scenario. Correspondingly, hydrogen production increases by 10.0\,TWh and utilization of electrolyzers drops by 3.4\% compared to the reference scenario.

On the supply side, differences between scenarios are more pronounced and thermal plants largely substitute imports. Compared to the reference scenario, thermal plants cover 78.9 instead of 25.2\,GW of the residual peak-load, while imports drop from 52.8 to 8.9\,GW. Total contribution to residual demand changes accordingly increasing from 8.8 to 36.2\,TWh for thermal plants but decreasing from 153.2 to 43.0\,TWh for imports. Similar to the reference scenario, thermal generation is almost equally divided between hydrogen turbines and gas engines.

\begin{figure}[!htbp]
	\centering
		\includegraphics[scale=0.13]{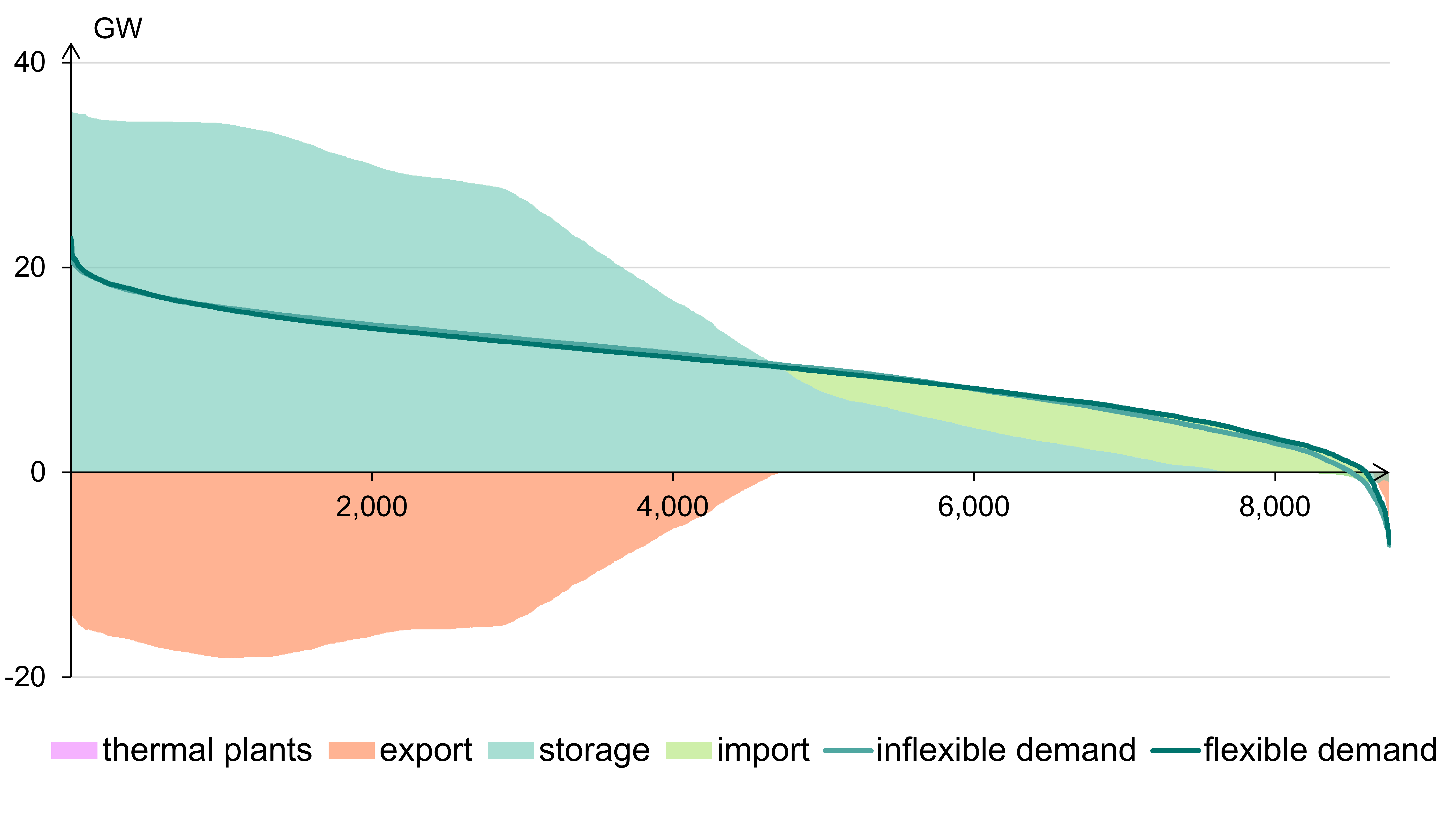}
	\caption{Residual load curves for Norway with grid expansion}
	\label{fig:dcGridNo}
\end{figure}

As a contrast to Germany, Fig.~\ref{fig:dcGridNo} shows residual load curves for Norway. Due to the exceptional Norwegian hydro resources, fluctuating renewables only supply 37.9\% of electricity, compared to 99.2\% in Germany. The model represents hydro reservoirs as storage systems with an exogenous inflow resulting in the large positive area in Fig.~\ref{fig:dcGridNo}. 

Due to its hydro reservoirs, Norway has a surplus of flexible generation and is even capable to export when residual demand is high. For the same reason, the country does not depend on flexible demand and its consideration hardly changes the residual load curve. Accordingly, Norway does not invest in heat or hydrogen storage, unlike Germany, and operates electrolyzers less flexible at a utilization rate of 84.4\%, greatly above the 44.4\% in Germany and close to the technical limit of 94.2\%. Residual load curves for other countries without exceptional hydro resources are similar to Germany and are provided in Fig.~\ref{fig:durOth} in appendix~\ref{c}.

\section{Conclusion} \label{5}

In this paper, we analyze flexible electrification in energy systems that rely on wind and solar as the main source of primary energy. Using a cost minimizing system model, we find substantial benefits of secondary demand from heating, transport, and industry adapting to fluctuating supply. In Germany, flexible demand halves the residual peak-load, halves the residual demand, and reduces excess generation by 80\%. Flexible operation of electrolyzers has the greatest impact and accounts for 42\% of the reduction in residual peak-load and 59\% in residual demand. District heating networks and BEVs provide substantial flexibility as well; the contribution of space and process heating is negligible. Leveraging this flexibility is cost-efficient but requires additional investments into systems for the storage and generation of hydrogen and heat.

To what extent flexible electrification is beneficial to reduce the residual load also depends on the availability of supply-side options to cover the residual load. Our analysis considers thermal plants, electricity storage, and the transmission grid. In the reference case, the latter is greatly expanded and covers most of the residual demand. Without grid expansion, the model deploys more thermal plants but does not substantially increase investment in demand-side flexibility. In case the supply-side already has a surplus of flexibility, there is no investment into demand-side flexibility at all. For instance, in Norway storage for hydrogen or heat is dispensable thanks to the large hydro reservoirs.

The purpose of our techno-economic analysis is not to assess the level of flexibility conceivable under current policy and market conditions. Instead, we identify where efforts to leverage the flexibility potential promise the greatest benefits. In light of our results, policy should prioritize the integration of electricity and hydrogen markets. Only if hydrogen production is sensitive to electricity prices, operators of electrolyzers have an incentive to adapt to renewable supply. The next priority is integrating district heating, followed by incentives for flexible charging of BEVs. Additional flexibility on the consumer level not only faces practical obstacles concerning privacy, automated control, and commercial aggregators but also has the smallest benefit on the system level.

For our analysis, it was key to consider a broad range of flexibility options and apply a high level of detail to spatio-temporal fluctuations of renewables. Nevertheless, future research can expand these qualities. Regarding flexibility, geothermal energy is a dispatchable technology to consider but excluded in this study due to a lack of data on regional potentials \citep{Ricks2022, MolarCruz2022}. Similarly, there is interest in carbon capture, utilization, and storage from a flexibility perspective \citep{Zantye2021}. Regarding detail, extending the analysis to cover multiple climatic years and include extreme weather conditions could improve the robustness of the results. The same applies to the representation of power grid constraints within market zones.

\section*{Supplementary material}

The model data and script is available on GitHub: \url{https://github.com/leonardgoeke/EuSysMod/releases/tag/flexibleElectrificationWorkingPaper}. 

The applied version of the AnyMOD.jl modeling framework is available here: \url{https://github.com/leonardgoeke/AnyMOD.jl/releases/tag/flexibleElectrificationWorkingPaper}

All files used to derive the model's quantitative inputs are shared on Zenodo \citep{dataSet}: \url{https://doi.org/10.5281/zenodo.6481534}

\section{Acknowledgments}

The research leading to these results has received funding from the European Union’s Horizon 2020 research and innovation program via the project "OSMOSE" under grant agreement No 773406 and from the German Federal Ministry for Economic Affairs and Energy via the project "MODEZEEN" (grant number FKZ 03EI1019D). A special thanks goes to all Julia developers.

\printcredits

\bibliographystyle{elsarticle-num-names}
\bibliography{cas-refs}

\appendix

\section{Detailed model structure} \label{a}

\begin{figure}[!htbp]
	\centering
		\includegraphics[scale=0.33]{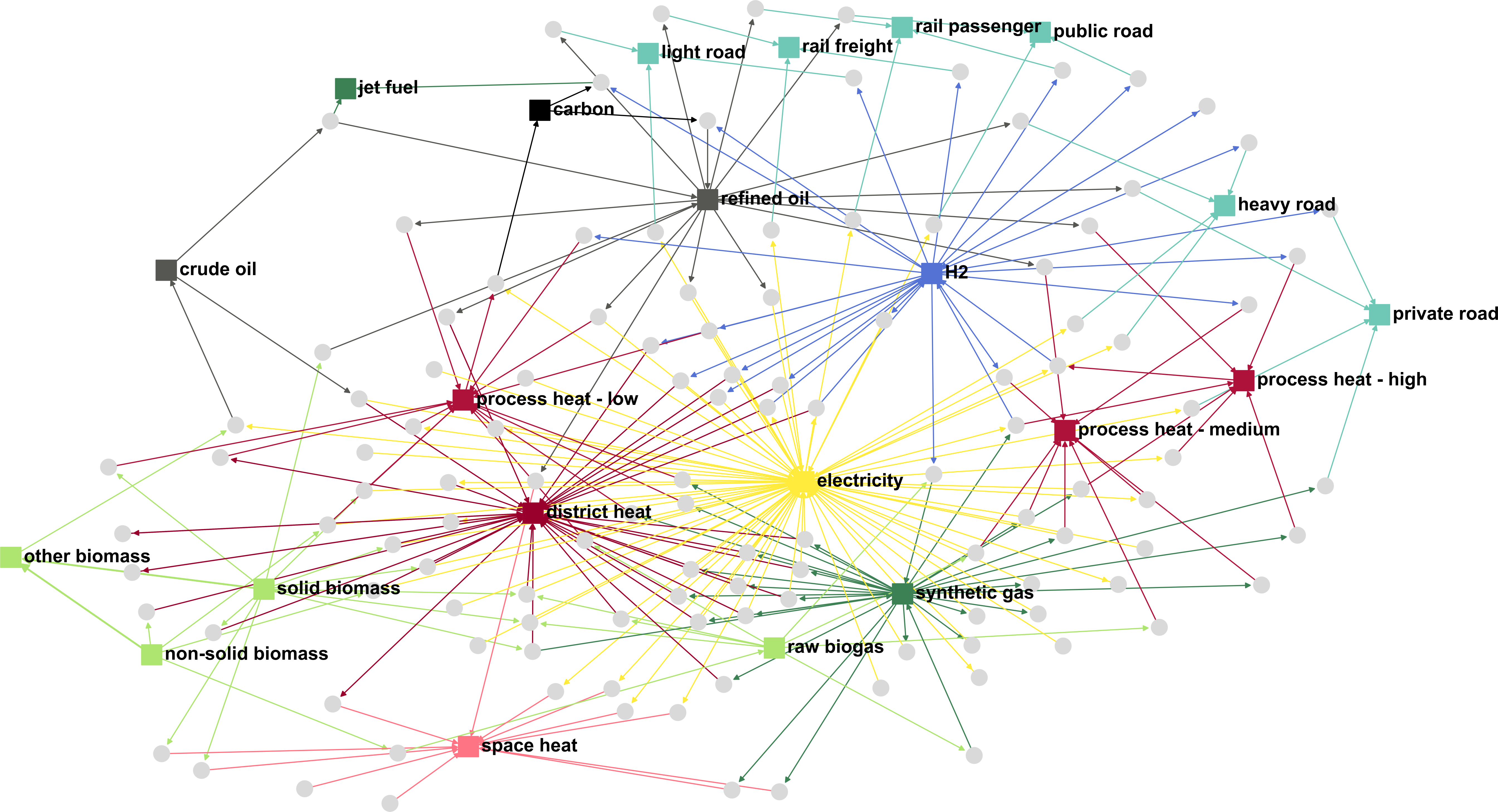}
	\caption{Full graph of model carriers and technologies}
	\label{fig:all}
\end{figure}

This section provides a comprehensive overview of all technologies and energy carriers considered in the model using graphs like Fig.~\ref{fig:powerAll} in the main part of the paper. In these graphs, vertices either represent energy carriers, depicted as colored squares, or technologies, depicted as gray circles. Entering edges of technologies refer to input carriers; outgoing edges refer to outputs. For illustrative purposes, Fig~\ref{fig:all} shows the graph of all carriers and technologies. In the following, subgraphs of this graph are used to go into further detail. 

\begin{figure}[!htbp]
	\centering
		\includegraphics[scale=0.33]{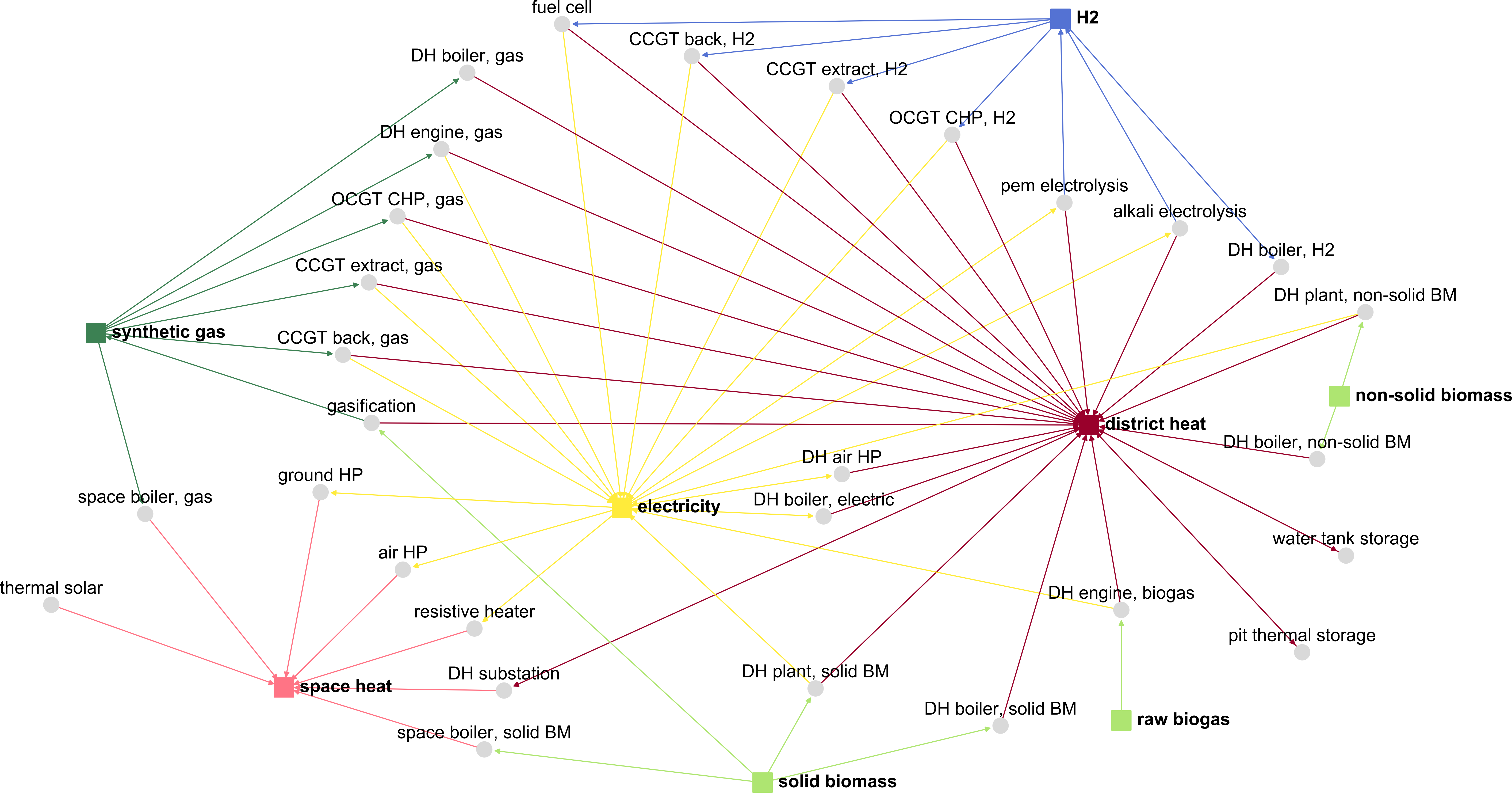}
	\caption{Subgraph for the supply of space and district heating}
	\label{fig:spaceDistrictHeat}
\end{figure}

Fig.~\ref{fig:spaceDistrictHeat} provides an overview of the technologies capable to supply space and district heat (DH) including their respective input carriers or additional outputs. Supply for district heat includes air heat pumps, different boilers, two distinct storage systems, and a broad range of combined-heat-power (CHP) technologies, some of which, like combined-cycle extraction turbines, can be operated flexibly and---within limits---increase their electricity output at the cost of reduced total efficiency. In addition, excess heat from electrolysis or gasifying solid biomass can be utilized as well. Demand for district heat is fully endogenous and induced by other technologies, for example, substations supplying space heat. 

\begin{figure}[!htbp]
	\centering
		\includegraphics[scale=0.33]{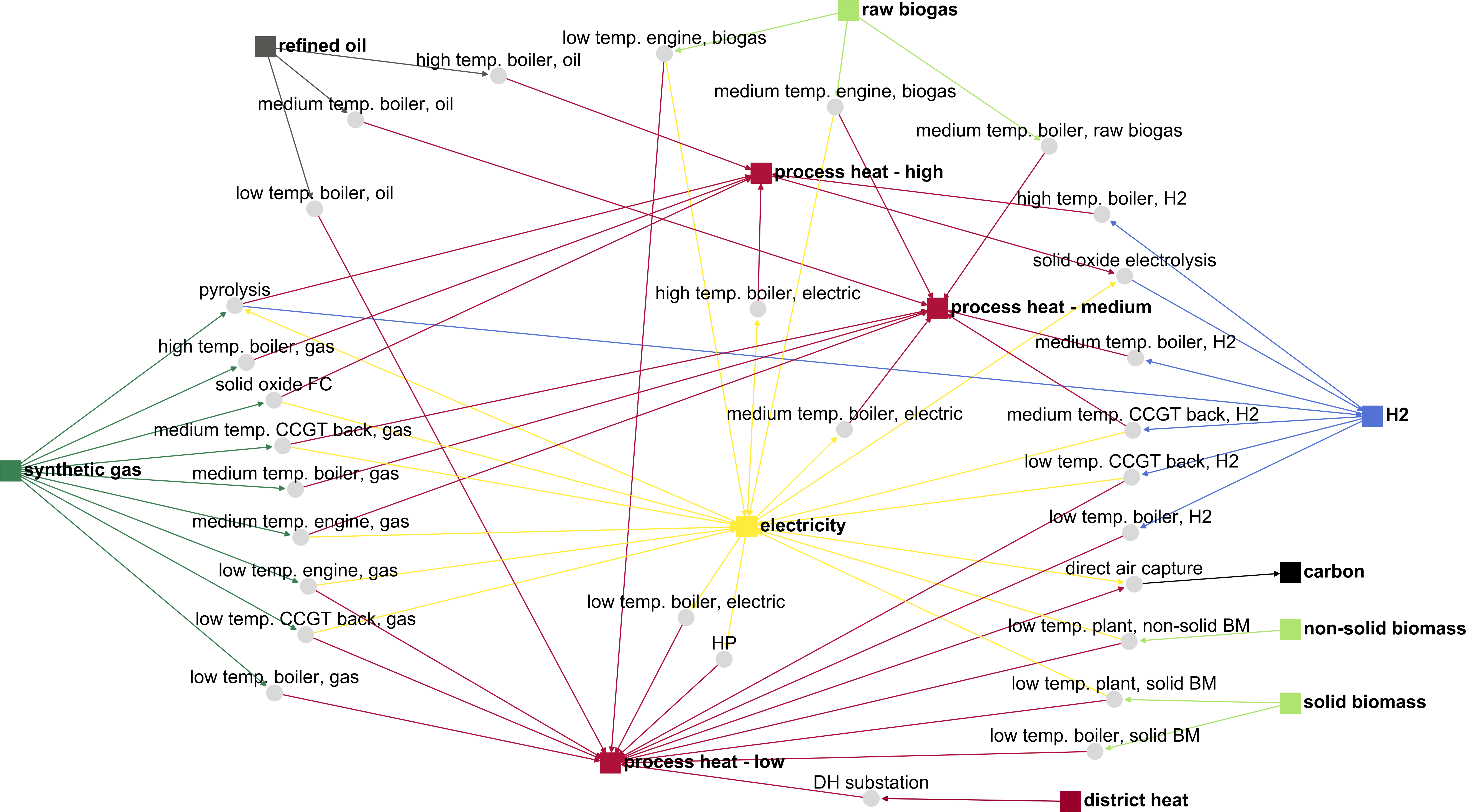}
	\caption{Subgraph for process heat}
	\label{fig:process}
\end{figure}

Fig.~\ref{fig:process} shows all technologies supplying heat for industrial processes, which are differentiated into three temperature levels. Low-temperature heat ranges up to 100\degree{}C, medium temperature from 100 to 500\degree{}C, and high temperature covers everything above 500\degree{}C. Heating requirements within the same temperature category can greatly vary by process and as a result, one technology might not be able to satisfy the entire demand of a certain category. For instance, not the entire heating demand above 500\degree{}C is eligible for electrification, because steel production in blast furnaces requires fuel combustion. To account for such limits, heat supply from district heating, engines, and combined-cycle gas turbines is limited to today's level as a conservative estimate.

\begin{figure}[!htbp]
	\centering
		\includegraphics[scale=0.33]{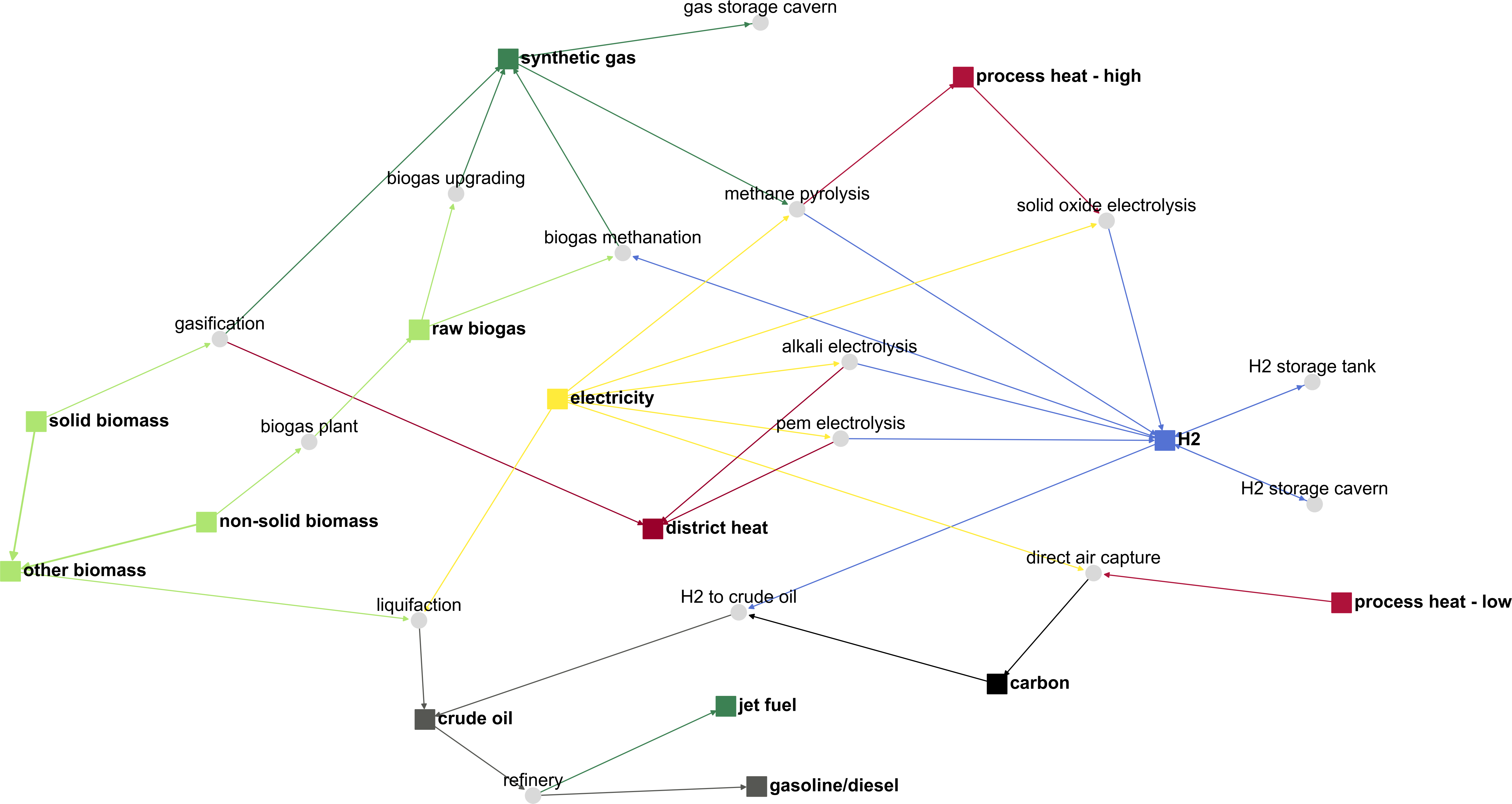}
	\caption{Subgraph for conversion of synthetic fuels}
	\label{fig:powerToX}
\end{figure}

Ultimately, Fig.~\ref{fig:powerToX} provides an overview of all technologies relevant for the generation and conversion of synthetic fuels. Thanks to the comprehensive sectoral perspective of our model, we are capable to consider the feed-in of excess heat from electrolysis or gasification into heating networks, but also the heat demand from processes like solid oxide electrolysis and direct air capture (DAC). For cavern storage of hydrogen regional capacity limits reflect that technology requires certain geological conditions and amount to 24.2\,PWh for Europe.   

\section{Model input data} \label{b}

Tables~\ref{tab:technology_data_elec} and \ref{tab:technology_data_stor} list the technology data for electricity generation in the system model from \citet{DEA}. For hydrogen fuelled plants, we assume a 15\% mark-up on the costs of the corresponding natural gas plant in line with \citet{Oberg2022}. This assumption neglects the cheaper alternative of retrofitting pre-existing gas plants instead. For all technologies, we assume a uniform interest rate of 5\%. The energy-to-power ratios for both battery technologies are constrained to be between 0.1 and 10. Tables~\ref{tab:final_demand} and \ref{tab:final_demand2} show the final demand for energy and transport services the model must meet.

\begin{table}[!htbp]
    \centering
    \caption{Technology data for electricity generation}
    \label{tab:technology_data_elec}
    \begin{tabular}{|l|c|c|c|c|c|}
        \hline        
        technology & investment costs & operational costs & lifetime & electrical  & availability  \\ 
        & [Mil. €/GW\textsubscript{in}] & [Mil. €/GW\textsubscript{in}n/a] & [a] & efficiency &  \\ \hline
        CC gas turbine, backpressure turbine & 586.0 & 13.7 & 25 & 50.3\% & 93.0\% \\ 
        CC gas turbine, extraction turbine & 480.7 & 15.9 & 25 & 58.3\% & 93.0\% \\ 
        CC H2 turbine, backpressure turbine & 673.9 & 13.7 & 25 & 50.3\% & 93.0\% \\ 
        CC H2 turbine, extraction turbine & 552.8 & 15.9 & 25 & 58.3\% & 93.0\% \\ 
        OC gas turbine & 177.8 & 3.2 & 25 & 41.5\% & 97.2\% \\ 
        OC gas turbine CHP & 226.6 & 7.7 & 25 & 41.5\% & 93.0\% \\ 
        OC H2 turbine & 204.4 & 3.2 & 25 & 41.5\% & 97.2\% \\ 
        OC H2 turbine CHP & 260.6 & 7.7 & 25 & 41.5\% & 93.0\% \\ 
        engine, biogas & 384.8 & 3.9 & 25 & 43.3\% & 95.1\% \\ 
        engine, diesel & 118.8 & 2.9 & 25 & 35.0\% & 96.8\% \\ 
        engine, gas & 221.5 & 2.9 & 25 & 47.6\% & 98.7\% \\ 
        engine CHP, gas & 415.5 & 4.2 & 25 & 46.9\% & 95.5\% \\
        non-solid biomass plant CHP & 272.4 & 9.7 & 25 & 28.1\% & 91.2\% \\
        solid biomass plant CHP & 870.5 & 24.9 & 25 & 26.8\% & 91.2\% \\ 
        polymer electrolyte fuel cell & 475.0 & 23.8 & 10 & 50.0\% & 99.7\% \\ 
        solid oxide fuel cell & 840.0 & 42.0 & 20 & 59.6\% & 100.0\% \\
        PV, openspace & 271.2 & 5.4 & 18 & ~ & ~ \\ 
        PV, rooftop residential & 693.8 & 9.5 & 40 & ~ & ~ \\
        PV, rooftop industry & 511.7 & 7.8 & 40 & ~ & ~ \\ 
        wind, onshore & 963.1 & 11.3 & 30 & ~ & ~ \\ 
        wind, offshore, shallow water & 1,577.6 & 32.5 & 30 & ~ & ~ \\ 
        wind, offshore, deep water & 1,777.3 & 32.5 & 30 & ~ &  \\ \hline
\end{tabular}
\end{table}

\begin{table}[!htbp]
    \centering
    \caption{Technology data for electricity storage}  
    \label{tab:technology_data_stor} 
    \begin{tabular}{|l|c|c|c|c|c|c|}
    \hline
        ~ & ~ & ~ & \multicolumn{2}{c|}{investment costs} & \multicolumn{2}{c|}{operational costs} \\
        technology & cycle efficiency & lifetime & power capacity & energy capacity & power capacity & energy capacity  \\ 
        ~ & ~ & [a] & [Mil. €/GW] & [Mil. €/GWh] & [Mil. €/GW/a] & [Mil. €/GWh/a] \\
        \hline
        lithium battery & 0.89 & 18 & 80.9 & 199.6 & 1.21 & 2.99 \\ 
        redox battery & 0.52 & 18 & 614 & 174.5 & 9.21 & 2.62 \\ 
        pumped storage & 0.81 & - & - & - & - & - \\ \hline
    \end{tabular}
\end{table}

\begin{table}[!htbp]
\footnotesize
    \centering
    \caption{Final demand for energy carriers} 
    \label{tab:final_demand2}
    \begin{tabular}{|l|c|c|c|c|c|c|c|c|c|c|c|c}
    \hline
    ~ & electricity & space heat & \multicolumn{3}{c|}{process heat [TWh]}  \\
    country &	 [TWh] &	 [TWh] & low (until 100\degree{}C) & medium (100 to 500\degree{}C) & high (above 500\degree{}C) \\ \hline
        Albania & 4.6 & 4.2 & 2.7 & 3.8 & 4.9 \\ 
        Austria & 62.8 & 45.2 & 17.0 & 26.2 & 20.7 \\
        Bosnia and Herzegovina  & 9.6 & 4.9 & 3.2 & 4.5 & 5.7 \\ 
        Belgium & 73.6 & 62.5 & 17.1 & 22.6 & 28.8  \\
        Bulgaria & 29.2 & 11.0 & 9.1 & 4.7 & 4.8  \\ 
        Switzerland & 51.8 & 43.7 & 7.4 & 30.8 & 6.6  \\
        Czech Republic & 57.8 & 38.9 & 17.5 & 16.7 & 19.0  \\ 
        Germany & 480.7 & 470.9 & 112.1 & 119.0 & 153.0 \\ 
        Denmark & 29.5 & 30.3 & 5.6 & 7.0 & 2.6  \\ 
        Estonia & 7.2 & 5.3 & 1.0 & 1.9 & 0.6  \\ 
        Spain & 241.9 & 62.8 & 25.5 & 57.2 & 54.1 \\ 
        Finland & 65.8 & 30.4 & 27.7 & 42.4 & 8.7  \\
        France & 390.3 & 257.8 & 40.4 & 57.1 & 64.6  \\ 
        Greece & 45.3 & 16.0 & 5.2 & 5.9 & 5.9  \\
        Croatia & 16.7 & 10.3 & 1.6 & 3.9 & 2.6  \\ 
        Hungary & 37.1 & 36.6 & 7.1 & 4.1 & 6.9  \\ 
        Ireland & 24.9 & 15.9 & 4.1 & 5.8 & 3.8  \\
        Italy & 260.6 & 215.9 & 57.1 & 48.5 & 72.7  \\ 
        Lithuania & 11.7 & 8.0 & 3.7 & 2.1 & 1.5  \\ 
        Luxembourg & 4.7 & 3.5 & 0.8 & 1.1 & 2.1  \\
        Latvia & 6.2 & 7.6 & 1.7 & 3.9 & 1.5  \\ 
        Montenegro & 2.8 & 0.9 & 0.5 & 0.7 & 1.1 \\
        Macedonia & 5.2 & 3.1 & 1.9 & 2.6 & 3.4  \\ 
        Netherlands & 107.7 & 62.8 & 30.0 & 28.1 & 42.5  \\ 
        Norway & 106.6 & 34.0 & 4.8 & 30.7 & 2.6 \\ 
        Poland & 145.1 & 98.0 & 23.6 & 40.8 & 42.5 \\ 
        Portugal & 46.5 & 9.1 & 8.4 & 16.1 & 9.1 \\ 
        Romania & 55.0 & 31.2 & 11.1 & 14.1 & 25.2 \\ 
        Serbia & 32.7 & 11.2 & 6.8 & 9.6 & 12.0  \\
        Sweden & 104.7 & 55.4 & 12.1 & 51.0 & 10.0 \\ 
        Slovenia & 12.3 & 6.5 & 1.7 & 2.6 & 2.3 \\ 
        Slovakia & 26.2 & 15.1 & 11.8 & 5.6 & 15.2 \\ 
        United Kingdom & 256.4 & 271.8 & 46.9 & 68.6 & 42.1 \\ \hline
    \end{tabular}
\end{table}

\begin{table}[!htbp]
\footnotesize
    \centering
    \caption{Final demand for transport services} 
    \label{tab:final_demand}
    \begin{tabular}{|l|c|c|c|c|c|c|c|c|c|c|c|c}
    \hline
    ~ & \multicolumn{3}{c|}{passenger transport [Gpkm]} & \multicolumn{3}{c|}{freight transport [Gpkm]} & ~ \\
    country & rail	& road private	& road public &	rail	& road heavy &	road light &	jet fuel [TWh] \\ \hline
        Albania &  0.1 & 8.6 & 3.0 & 2.5 & 3.8 & 0.0 & 0.0 \\ 
        Austria &  14.0 & 83.8 & 11.0 & 12.6 & 25.6 & 0.9 & 18.1 \\
        Bosnia and Herzegovina  & 0.1 & 11.5 & 4.0 & 3.4 & 5.1 & 0.1 & 0.0 \\ 
        Belgium & 11.7 & 117.8 & 14.9 & 5.7 & 30.9 & 3.9 & 51.7 \\
        Bulgaria &  1.4 & 55.9 & 7.8 & 7.1 & 20.2 & 0.3 & 5.4 \\ 
        Switzerland & 20.8 & 80.6 & 6.4 & 6.4 & 11.6 & 0.3 & 0.5 \\
        Czech Republic &  10.0 & 76.5 & 17.7 & 19.2 & 46.6 & 3.8 & 5.0 \\ 
        Germany & 101.7 & 950.7 & 64.8 & 84.8 & 303.3 & 8.6 & 151.0 \\ 
        Denmark &  6.9 & 67.1 & 7.9 & 2.0 & 14.8 & 0.2 & 4.3 \\ 
        Estonia & 0.4 & 11.3 & 2.4 & 4.1 & 4.7 & 0.1 & 0.0 \\ 
        Spain & 31.0 & 370.9 & 35.0 & 11.4 & 213.9 & 3.1 & 6.1 \\ 
        Finland & 4.6 & 67.8 & 8.1 & 11.8 & 28.4 & 0.5 & 16.7 \\
        France & 95.3 & 770.4 & 59.2 & 19.5 & 169.3 & 4.8 & 114.2 \\ 
        Greece &  1.2 & 106.0 & 21.0 & 0.6 & 27.0 & 1.2 & 69.5 \\
        Croatia & 0.8 & 25.6 & 5.3 & 3.6 & 12.2 & 0.3 & 3.3 \\ 
        Hungary &  6.7 & 55.1 & 16.2 & 14.5 & 35.8 & 1.1 & 6.1 \\ 
        Ireland &  1.1 & 28.5 & 5.2 & 0.1 & 9.0 & 0.2 & 0.0 \\
        Italy &  59.0 & 768.2 & 89.9 & 17.0 & 96.3 & 16.4 & 66.9 \\ 
        Lithuania &  0.4 & 30.3 & 2.9 & 112.4 & 52.3 & 0.8 & 29.6 \\ 
        Luxembourg &  0.4 & 7.2 & 1.1 & 0.7 & 7.3 & 0.1 & 0.0 \\
        Latvia &  0.6 & 15.3 & 2.6 & 46.9 & 14.8 & 0.2 & 0.0 \\ 
        Montenegro & 0.1 & 4.1 & 0.1 & 0.9 & 1.4 & 0.0 & 0.0 \\
        Macedonia &  0.1 & 7.5 & 2.2 & 2.1 & 3.2 & 0.0 & 0.0 \\ 
        Netherlands &  24.5 & 187.4 & 6.8 & 8.7 & 66.7 & 1.6 & 215.6 \\ 
        Norway & 3.6 & 64.6 & 4.2 & 3.9 & 20.7 & 0.6 & 17.0 \\ 
        Poland & 21.8 & 218.7 & 35.6 & 127.9 & 340.0 & 9.0 & 35.1 \\ 
        Portugal & 4.0 & 85.2 & 7.0 & 5.1 & 30.4 & 0.6 & 29.1 \\ 
        Romania & 4.4 & 81.4 & 15.6 & 40.1 & 60.3 & 0.8 & 11.6 \\ 
        Serbia &  0.3 & 28.8 & 10.0 & 8.5 & 12.7 & 0.2 & 0.0 \\
        Sweden &  11.7 & 100.0 & 8.7 & 19.2 & 42.0 & 0.6 & 7.7 \\ 
        Slovenia &  0.5 & 26.3 & 3.6 & 13.1 & 23.7 & 0.3 & 0.0 \\ 
        Slovakia & 3.7 & 27.9 & 6.1 & 17.2 & 32.5 & 1.5 & 2.5 \\ 
        United Kingdom & 69.1 & 670.3 & 36.5 & 16.7 & 153.7 & 7.1 & 143.9 \\ \hline
    \end{tabular}
\end{table}

\section{Additional model results} \label{c} 

\begin{figure}[!htbp]
	\centering
		\includegraphics[scale=0.13]{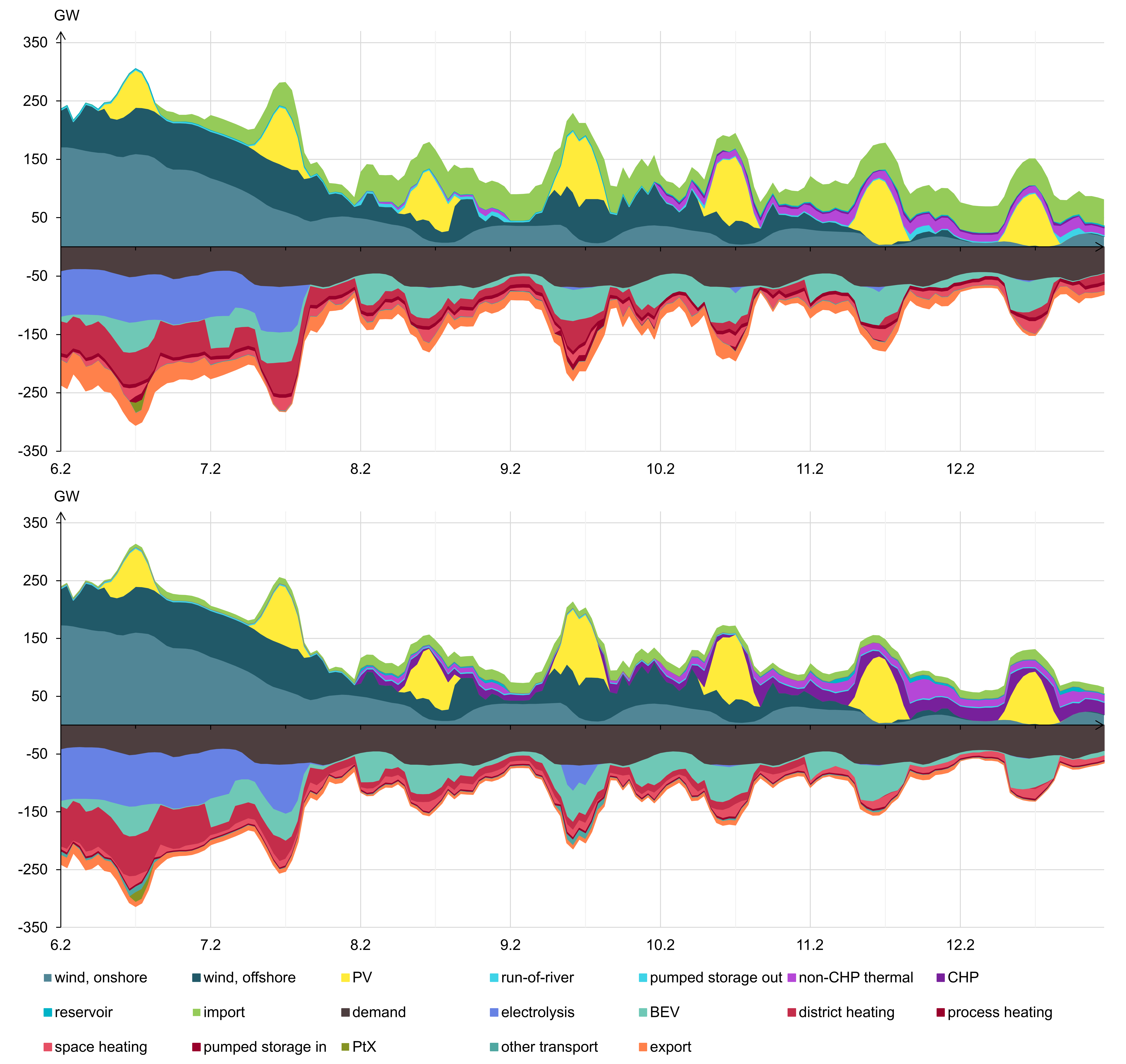}
	\caption{Supply and demand in Germany for one week}
	\label{fig:ts}
\end{figure}

Fig.~\ref{fig:ts} shows hourly electricity supply and demand for an exemplary February week in Germany. 

\begin{sidewaysfigure}
	\centering
		\includegraphics[scale=0.5]{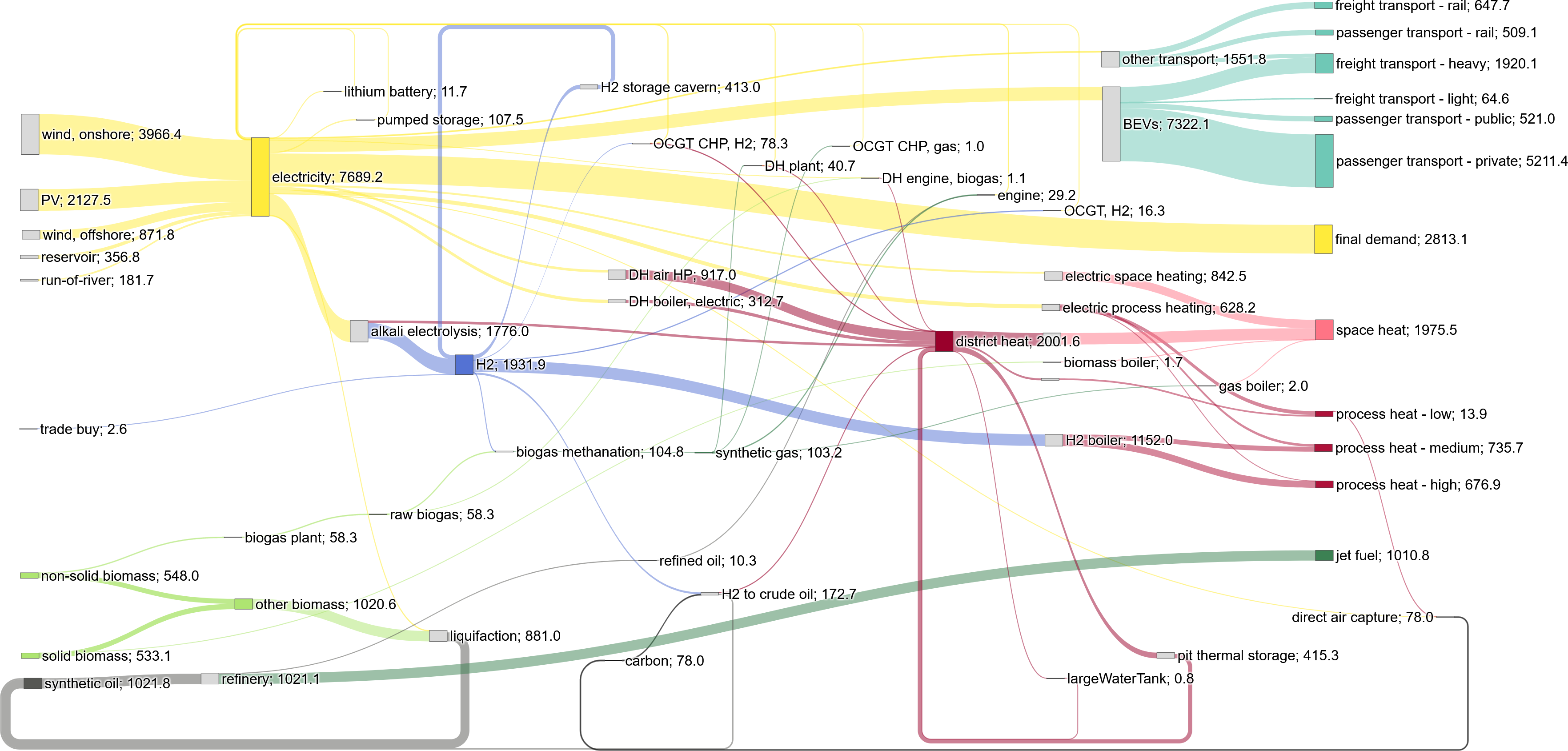}
	\caption{Sankey diagram without grid expansion, in TWh/Gpkm/Gtkm}
	\label{fig:sankeyGridNoGrid}
\end{sidewaysfigure}

\begin{figure}[!htbp]
	\centering
		\includegraphics[scale=0.85]{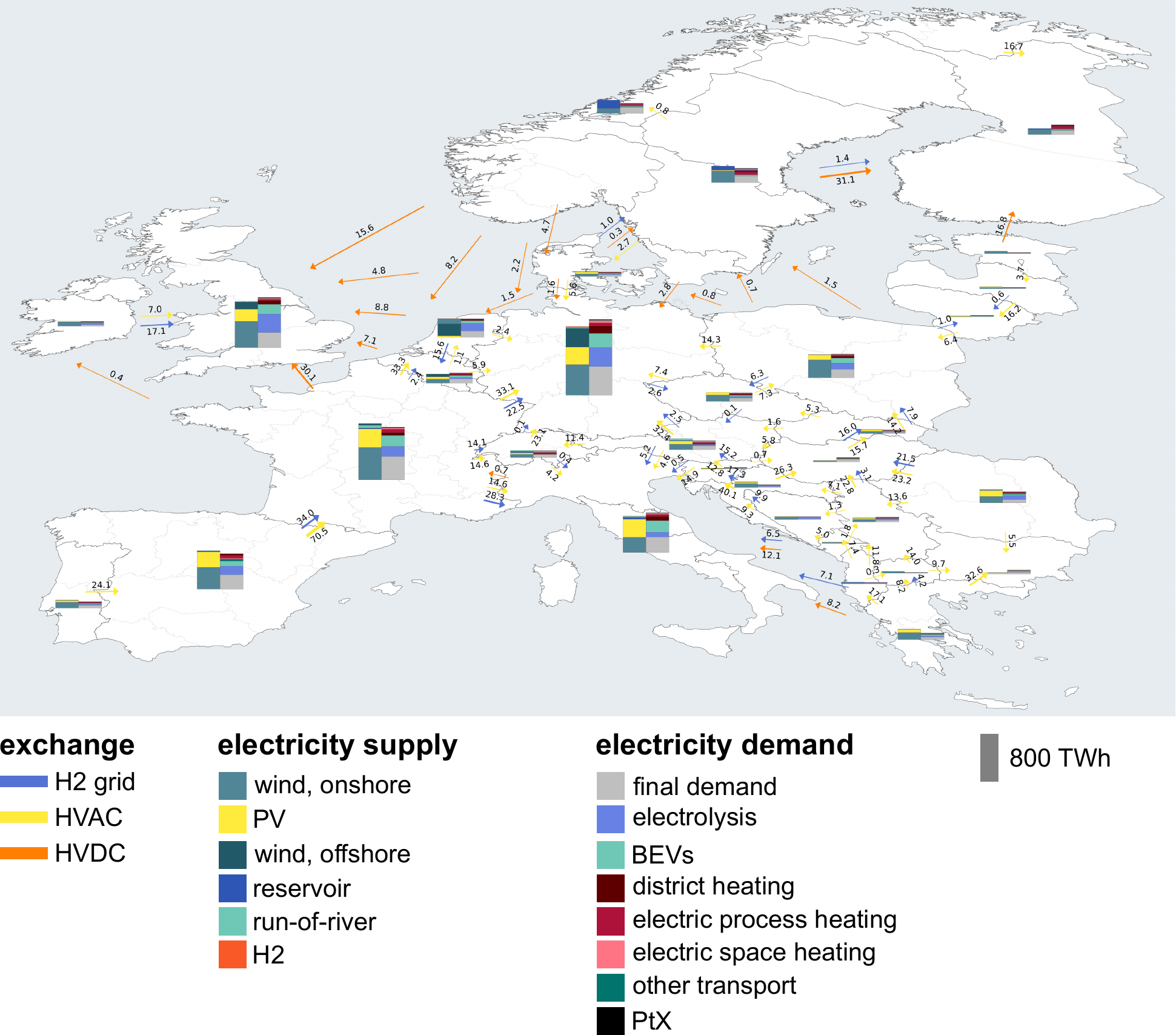}
	\caption{Electricity generation and exchange without grid expansion}
	\label{fig:resMapNoGrid}
\end{figure}

Analogously to section~\ref{enBal} describing results for the reference scenario with grid expansion, Figs.~\ref{fig:sankeyGridNoGrid} and \ref{fig:resMapNoGrid} show the Sankey diagram and map for the scenario without grid expansion.

\begin{figure}[!htbp]
	\centering
		\includegraphics[scale=0.13]{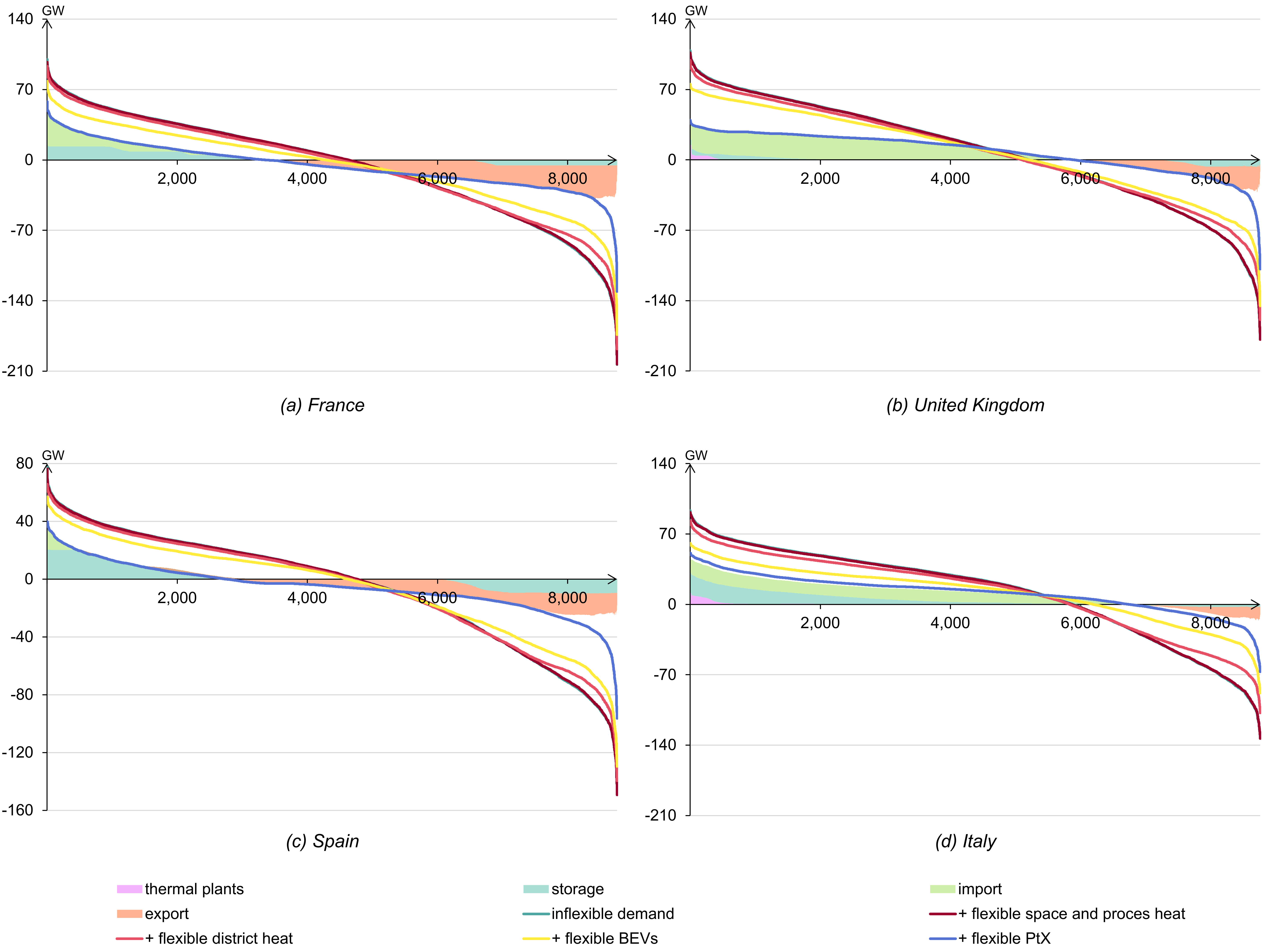}
	\caption{Residual load curves with grid expansion}
	\label{fig:durOth}
\end{figure}

\end{document}